\documentclass[11pt, letterpaper]{article}

\usepackage{graphicx}
\usepackage[square,comma,numbers]{natbib}
\usepackage{amsmath}
\usepackage{amssymb}
\usepackage{amsmath}
\usepackage{amsfonts}
\usepackage{mathtools}
\usepackage[super]{nth}
\usepackage{setspace}
\usepackage{enumitem} 
\usepackage{epstopdf}
\usepackage{color}
\usepackage[letterpaper,left=1in,right=1in,top=1in,bottom=1in]{geometry}
\usepackage[]{algorithm2e}
\usepackage{multirow}
\usepackage{longtable}
\usepackage{rotating}
\usepackage{mathrsfs}
\usepackage{subfigure}
\usepackage{float}
\usepackage{graphicx}
\usepackage{caption}
\usepackage{bm}
\usepackage{algorithmic}

\newtheorem{remark}{Remark}

\newtheorem{theorem}{Theorem}

\newtheorem{definition}{Definition}

\newcommand\norm[1]{\left\lVert#1\right\rVert}

\DeclareMathOperator*{\argmax}{argmax}

\graphicspath{{Figures//}}

\title{\Large \bf Control invariant set enhanced {safe} reinforcement learning: improved sampling efficiency, guaranteed stability and robustness}
\author{
	\centerline{\normalsize Song Bo$^{a}$, Bernard T. Agyeman$^{a}$, Xunyuan Yin$^{b}$, Jinfeng Liu$^{a,}$\thanks{Corresponding author: J. Liu. Tel: +1-780-492-1317. Fax: +1-780-492-2881. Email: jinfeng@ualberta.ca.}}
	\vspace{5mm}\\
	\centerline{\small $^{a}$Department of Chemical \& Materials Engineering, University of Alberta,}\\
	\centerline{\small Edmonton, AB, Canada T6G 1H9.}\\
	\centerline{\small $^{b}$School of Chemistry, Chemical Engineering and Biotechnology, Nanyang Technological University,}\\
	\centerline{\small Singapore 637459.}\\}
\allowdisplaybreaks

\begin{document}
	\date{}
	\maketitle
	%\doublespacing
	%\onehalfspacing
	\setstretch{1.5}
	{}
	\begin{abstract}
	Reinforcement learning (RL) is an area of significant research interest, and safe RL in particular is attracting attention due to its ability to handle safety-driven constraints that are crucial for real-world applications. This work proposes a novel approach to RL training, called control invariant set (CIS) enhanced RL, which leverages the advantages of utilizing {the explicit form of} CIS to improve stability guarantees and sampling efficiency. {Furthermore, the robustness of the proposed approach is investigated in the presence of uncertainty.} The approach consists of two learning stages: offline and online. 
	In the offline stage, CIS is incorporated into the reward design, initial state sampling, and state reset procedures. {This incorporation of CIS facilitates improved sampling efficiency during the offline training process.} In the online stage, RL is retrained whenever the predicted next step state is outside of the CIS, which serves as a stability criterion, {by introducing a Safety Supervisor to examine the safety of the action and make necessary corrections. The stability analysis is conducted for both cases, with and without uncertainty.} 
	To evaluate the proposed approach, we apply it to a simulated chemical reactor. The results show a significant improvement in sampling efficiency during offline training and closed-loop stability guarantee in the online implementation, with and without uncertainty.
	\end{abstract}
	\clearpage

\section{Introduction}

In process control, model predictive control (MPC) is a standard approach to optimal control. It is formulated as a constraint optimization problem in which the safety constraints are taken into account explicitly \cite{mayne2000constrained}. However, for large-scale systems, MPC may suffer from high computational complexity. Reinforcement learning (RL), as one main component of machine learning, provides an alternative to MPC for optimal control and can shift the complex optimization calculations to offline training based on a model \cite{sutton2018reinforcement}. It has gain significant attention in different industries, (for example, game \cite{mnih2015human}, finance \cite{singh2022reinforcement}, energy \cite{chen2022reinforcement}) for decision-making and control purposes. 

RL is a class of optimal control algorithms that enables machines to learn an optimal policy (closed-loop control law), by maximizing future rewards through repetitive interactions with the environment~\cite{sutton2018reinforcement}. It uses a trial-and-error approach to interact with the environment, allowing it to learn and find the optimal policy even in the absence of prior knowledge of the process. In addition, the consideration of future rewards in RL ensures that current decisions are beneficial in the long run. However, the standard RL approach does not incorporate safety constraints in its design and does not guarantee closed-loop stability, which limits its use in real-world applications \cite{garcia2015comprehensive}. To address these challenges, safe RL algorithms have been developed, which explicitly consider safety constraints during training and ensure closed-loop stability in the learned policy.

Safe RL is a class of RL algorithms that aims to achieve a stable closed-loop control system and satisfy safety constraints. Different approaches have been reviewed in the literature \cite{garcia2015comprehensive, osinenko2022reinforcement, gu2022review}. One common approach is to formulate the problem as a constrained Markov decision process (CMDP), where a cost function, defined in terms of safety constraints, is used to constrain undesired actions during policy optimization \cite{kadota2006discounted, chow2017risk}. This transformation turns the original problem of maximizing returns into a new problem that requires balancing both return maximization and constraint violation minimization. Another approach is to integrate safety constraints into the exploration process by modifying the value function, encouraging the agent to choose safer actions during exploration \cite{law2005risk, gehring2013smart}. However, these approaches do not provide guaranteed safety.
Another strategy is to combine MPC with RL by treating MPC as parameterized value or policy neural networks \cite{zanon2019practical, gros2021reinforcement}. {Although safety constraints can be explicitly handled in MPC,} these approaches still involve solving the MPC optimization problem recursively, leading to significant computational burden.

Sampling efficiency is {another critical issue that has limited real-world applications of RL} \cite{yu2018towards}. {While several works in the field of safe RL have addressed safety considerations, {they have also recognized the importance of sampling efficiency as a by-product of their approaches} \cite{law2005risk, ma2021model}.} In the context of stability, an RL algorithm {could require}  a significant number of agent-environment interactions to achieve a stable and optimal control policy, leading to prohibitively high costs in its application. To address this challenge, an intuitive solution is to {constrain} the agent to interact only with the controllable states of the environment. By doing so, the agent can {learn} a policy that {confines} the system within the controllable state space, achieving the stability guarantee while eliminating unnecessary interactions with uncontrollable states and improving sampling efficiency. Unfortunately, {such environments where only controllable states are present are generally not available in practice.}

In the field of control theory, it is widely acknowledged that CIS plays a crucial role in ensuring the stability of control systems \cite{decardi2022robust}. These sets characterize the states for which a feedback control law is always available to maintain the system within the set~\cite{blanchini1999set}. Incorporating the concept of CIS in RL is expected to improve the sampling efficiency and stability guarantee by constraining the agent's interactions to controllable states. {This approach enables the agent to learn a policy that maintains the system within the CIS using fewer interactions with the environment, thereby achieving stability. Notably, the concept of CIS has been adopted in RL designs to achieve closed-loop stability.} 
{Researchers have explored two main approaches in this context. The first approach assumes that the CIS is unknown and is simultaneously approximated along with the policy. This approach aims to constrain the policy based on the estimated CIS to achieve safety, but the safety guarantee may be compromised if the CIS approximation is inaccurate. This approach aligns with the concept of CMDP where the safety constraint is explicitly defined using CIS concepts \cite{ma2021model}. The second approach assumes knowledge of the CIS information and involves filtering or projecting risky actions into safe ones, typically by incorporating a standalone safety filter after the RL policy \cite{alshiekh2018safe, gros2020safe, li2020robust, tabas2022computationally}. However, since the filter only considers safety, it does not always preserve the optimality that the controller aims to achieve. To address this, \cite{gros2020safe} proposes embedding CIS in the last layer of the RL policy network, enabling back-to-back training to achieve both safety and optimality.}
Due to the challenge of obtaining CIS for general nonlinear systems, aforementioned works have shifted their focus towards implicit methods that utilize control barrier functions (CBF), Hamilton-Jacobi (HJ) reachability analysis, and safe backup controllers to define {CIS} and design filters indirectly \cite{brunke2022safe}. {A comprehensive introduction and comparison between CBF and HJ is presented in \cite{li2021comparison}.}
It is worth noting that the studies combining CIS and RL have primarily focused on robotics, with limited research conducted in process control, {in which} process systems are in general highly nonlinear, tightly interconnected and of large-scale. These features pose challenges for applying CBF- and HJ-based algorithms mentioned above. Additionally, process control often involves control objectives beyond set-point or reference trajectory tracking, such as zone tracking \cite{decardi2022robust, zhang2020} and economic optimization \cite{ellis2014}. These objectives introduce additional complexities that make the aforementioned approaches less suitable.

While the construction of a CIS is not a trivial task, various methods have been developed in the past decade. For example, algorithms for constructing or approximating the CIS for constrained linear systems \cite{rungger2017computing, rakovic2005invariant} and general nonlinear systems \cite{homer2017constrained, fiacchini2010computation} have been proposed. Graph-based approaches to find the outer and inner approximations of robust CIS (RCIS) of nonlinear systems has also been developed \cite{decardi2021computing}. Over the past couple years, data-driven approaches have also been used to find the invariant sets of nonlinear systems, which approximate invariant sets using neural nets \cite{chen2021learning, bonzanini2022scalable}. {These approaches have paved the way for studying safe RL that explicitly incorporates a CIS. In such concept, the CIS serves as a safe state space containing only controllable states that the RL agent can explore and exploit.}

The above considerations motivate us to study the explicit integration of RL and CIS for process control, where the CIS can serve as a state space for the RL agent to explore, safely. {Because the work focuses on how RL agent interacts with the safe state space, CIS, it requires minimal modification to the RL agents themselves and the economic or zone tracking objectives which are common in process control can be incorporated in the reward function design naturally. Further, the sampling efficiency, the optimality with respect to different control objectives and the robustness are kept in mind while designing the CIS-enhanced RL method.}
Specifically, in this work, the CIS of a nonlinear process is assumed to be available. Then, a two-stage CIS enhanced RL is proposed to improve the sampling efficiency and guarantee the stability. The first stage involves offline training with a process model and the CIS. Due to the potential disastrous consequences of failed process control, the use of a model to pre-train the RL offline can provide a significant amount of data with strong temporal correlation and broad coverage of various scenarios. The introduction of CIS has the potential to narrow down the state space, reduce the training dataset size, and provide guidance on agent exploration. However, exhaustive training cannot guarantee that the RL agent has encountered every scenario, which may result in instability in online implementation. Hence, the second online implementation stage involves online learning when the safety constraint is violated. A new control implementation strategy{, named Safety Supervisor,} is proposed to ensure closed-loop stability. {Furthermore, the robustness of the proposed approach is studied in the presence of uncertainty. The stability proofs are provided for both deterministic cases and scenarios with uncertainty. Moreover, the algorithm is adapted to accommodate different control objectives, allowing us to achieve both control performance and stability simultaneously. Finally}, the proposed approach is applied to a chemical reactor to demonstrate its applicability and efficiency.

This paper builds upon the preliminary findings reported in \cite{bo2023control}. In this work, we provide more comprehensive explanations and extensive simulation results for the deterministic case. Additionally, we introduce the algorithm design and present simulation results for the uncertain case. Furthermore, we provide formal stability proofs for the proposed approach.

%\section{Notation}
%{\color{red}In this manuscript, we use the symbol `$\setminus$' to denote set subtraction, which is defined as follows: $\mathbb{A} \setminus \mathbb{B} \coloneqq \{x\in \mathbb{A} \mid x \notin \mathbb{B}\}$. In addition, bold face variables are used throughout the manuscript to represent sets. For example, the bold face variable $\bm{u}$ represents a set that contains $u$ scalar variables.}

\section{Preliminaries}

\subsection{System description}

In this work, we consider a class of nonlinear processes that can be described by the following discrete-time state-space form:
\begin{equation} \label{eqn:nonlinear}
	x_{k+1}=f(x_k, u_k,{ w_k})
\end{equation}
where $x \in X \subseteq \mathbb{R}^{n}$ and $u \in U \subseteq \mathbb{R}^{m}$ denote the state and the input {vectors} of the process with $X$ and $U$ being the physical constraints on $x$ and $u$. {The variable $w \in W \subseteq \mathbb{R}^n$ represents the disturbance vector and is bounded by the a known set $W$.} {The function }$f:~\mathbb{R}^{n}~\times~\mathbb{R}^{m}~\times~\mathbb{R}^{n}~\rightarrow~\mathbb{R}^{n}$ is a nonlinear function mapping the present state to the state at the next time instant, $k$ represents the time index.

\subsection{Reinforcement learning}

RL broadly represents {the class of }data-driven learning algorithms in which an agent is trying to learn a closed-loop policy $\pi(u|x)$, a conditional probability of prescribing $u$ at given state $x$, by interacting with the environment. 
The Markov decision process (MDP) is utilized to formulate the environment. The environment receives the action prescribed by the agent and provides the resulting reward and state of the system back to the RL agent. {Hence, one example of the data tuple required by RL is shown as follows:
	\begin{equation}\label{eqn:data_tuple}
		(x_k, u_k, r_{k+1}, x_{k+1})
	\end{equation}
} 
and the state transition dynamics of the MDP is shown below:
\begin{equation} \label{eqn:state_transition}
	{P}(r_{k+1}, x_{k+1}|x_k, u_k)
\end{equation}
where $x_k$ denotes the current state of the environment, $u_k=\pi(x_k)$ is the action prescribed by the agent based on the learned policy, $r_{k+1}$ represents the reward used for criticizing the action, $x_{k+1}$ represents the state sampled at the next sampling time instant, {${P}$} denotes the conditional probability of the state transition.

As in \cite{sutton2018reinforcement}, the RL problem can be formulated as the following:
\begin{equation}
	\pi^* = \argmax_{\pi} \quad {\mathbb{E}_{\pi}[G_k|x_k, u_k]}
\end{equation}
where $G_k$ denotes the return accumulating the reward $r$ in long run. The optimal policy $\pi^*$ is found when the expected return following the such policy is maximized.

In this work, the environment dynamics that describe the transition from $x_k$ to $x_{k+1}$ is represented by the nonlinear system of equation~\eqref{eqn:nonlinear}.

\subsection{Control invariant sets}

A control invariant set of a system is a set of states in which the system can stay inside all the time by following a feedback control law. The definition of the control invariant set is given below:
\begin{definition}[c.f. \cite{blanchini1999set}]\label{def:cis}
	{Without uncertainties, }the set $R \subseteq X$ is said to be a control invariant set for system~\eqref{eqn:nonlinear} if for any $x_k\in R$, there exists an input $u_k\in U$ such that $x_{k+1}\in R$.
\end{definition}

{Furthermore, the RCIS is defined below with the presence of the uncertainty:
	\begin{definition} [c.f. \cite{blanchini1999set}]\label{def:rcis}
		The set $R_r \subseteq X$ is said to be a RCIS for system~\eqref{eqn:nonlinear} if it is a CIS for all $w_k \in W$.
\end{definition}}

In the control literature, {(R)}CISs play an important role in ensuring the stability of the closed-loop systems. For example, {(R)}CISs are commonly {used} in MPC designs as a terminal constraint for achieving guaranteed stability and feasibility \cite{mayne2000constrained,decardi2022robust,cannon2003nonlinear}. % In \cite{decardi2022robust}, an economic zone designed within the CIS ensures the stability and optimizes the economic objective simultaneously. Decardi-Nelson and Liu \cite{decardi2021computing} proposed a graph-based approach to compute the outer and inner approximations of CIS of a discrete-time nonlinear system in the form of equation~\eqref{eqn:nonlinear}. 

{A common way to describe a CIS is to represent it as a polytope \cite{rungger2017computing, decardi2021computing}. In this work, we also assume that the CIS is represented as a polytope as follows:
	\begin{equation}\label{eqn:polytope}
		R=\{x \in \mathbb{R}^n:~Ax-b\leq \textbf{0}\}
	\end{equation}
	where the matrix $A \in \mathbb{R}^{c\times n}$ and the vector $b \in \mathbb{R}^{c}$ define the hyperplanes bounding the polytope, comprising a total of $c$ affine constraints. The vector $\textbf{0} \in \mathbb{R}^c$ has all its elements set to 0.}

\begin{remark}
	The distinction between CIS definition~\eqref{eqn:polytope} and HJ- or CBF-based safe region definition is that the safety constraint in~\eqref{eqn:polytope} is capable to directly assess whether a state is control invariant, whereas the other two require the computation of a value function $V(x)$. Taking HJ as an example, a value function is defined to find the best action to fight against the worst-case scenario in an infinite horizon \cite{brunke2022safe}:
	\begin{equation}\label{eqn:HJ}
		V(x) = \max_{u_k\in U}\min_{w_k\in W}\inf_{k\geq 0}h(f(x_k,u_k,w_k))
	\end{equation}
	where the safety constraint $h(\cdot): \mathbb{R}^n\rightarrow \mathbb{R}$, deemed safe when $h(\cdot)\geq 0$, is less strict than the constraint in~\eqref{eqn:polytope} so that it does not guarantee a control-invariant state when satisfied. The two inner $\min$ operators search for the worst disturbance $w_k$ that can lead to a violation of the constraint, $h(\cdot)<0$, over an infinite time horizon. The outer $\max$ operator then identifies the best action to prevent constraint violation.
	
	The safe region, based on HJ reachability analysis, may be defined as follows:
	\begin{equation}
		S=\{x\in \mathbb{R}^{n}: V(x)\geq 0\}
	\end{equation}
	where a state $x$ belongs to $S$ if the corresponding value function $V(x)$ is non-negative, and vice versa.
	While the safety constraint $h(\cdot)$ is straightforward to define using HJ approach, the challenge lies in solving equation~\eqref{eqn:HJ}, since it is only computationally tractable for specific model structures, such as linear or control-affine processes, or low-dimension nonlinear systems \cite{li2021comparison}. %In this work, by directly utilizing a stringent safety constraint in equation~\eqref{eqn:polytope}, the step of finding the value function is skipped. This assumes that the explicit form of the CIS, equation~\eqref{eqn:polytope}, is known for the studied nonlinear systems.
\end{remark}

\subsection{Problem formulation}
Standard RL does not consider the safety constraints which obstructs its application{. Also, conventional RL typically requires} a significant amount of data for training. CISs inherently provide the region of operation that is stable and may be integrated in RL offline and online training to ensure closed-loop stability. In addition, the introduction of the CIS is possible to narrow down the state space, to reduce the training dataset size, and to provide guidance on agent exploration.

The objective of this work is to propose a CIS enhanced RL and its training method to guarantee the closed-loop stability and improve the sampling efficiency during RL training, by incorporating the CIS knowledge into RL. The RL optimization can be described as follows:
\begin{equation}
	\begin{aligned}
		\pi^* = \argmax_{\pi} \quad & {\mathbb{E}_{\pi}[G_k|x_k, u_k]}\\
		\textrm{s.t.} \quad & x_{k}\in R\\
		& u_k\in U
	\end{aligned}
\end{equation}
where the state constraint as well as the input constraint are considered. 

\section{Proposed approach{ - deterministic case}} \label{sec:proposed_procedure}

In this section, we present the proposed CIS enhanced RL {without taking into account model uncertainties}. In the proposed approach, it includes both offline training and online training. The first step is to train the RL with the CIS information offline to achieve a near-optimal policy. The incorporation of the CIS in the offline training can significantly improve its sampling efficiency {since the amount of data needed for training is reduced; this} will be demonstrated in the simulations. While the CIS is used in offline training, the offline trained policy does not guarantee the closed-loop stability. In order to ensure the closed-loop stability, the RL should further be trained during its online implementation. A new control implementation strategy is proposed to ensure that the applied control actions ensure the closed-loop stability. Figure~\ref{fig:proposed_alg} illustrates the proposed approach {and the difference between the proposed approach and the standard RL are highlighted in blue. {To distinguish the responses of the Environment and the mathematical model, $x$ and $\hat{x}$ denote the actual response of the process and the predicted state of the model, respectively. }The details of the two steps are explained below. 
	
	\begin{figure}
		\centering
		\includegraphics[width=0.59\textwidth]{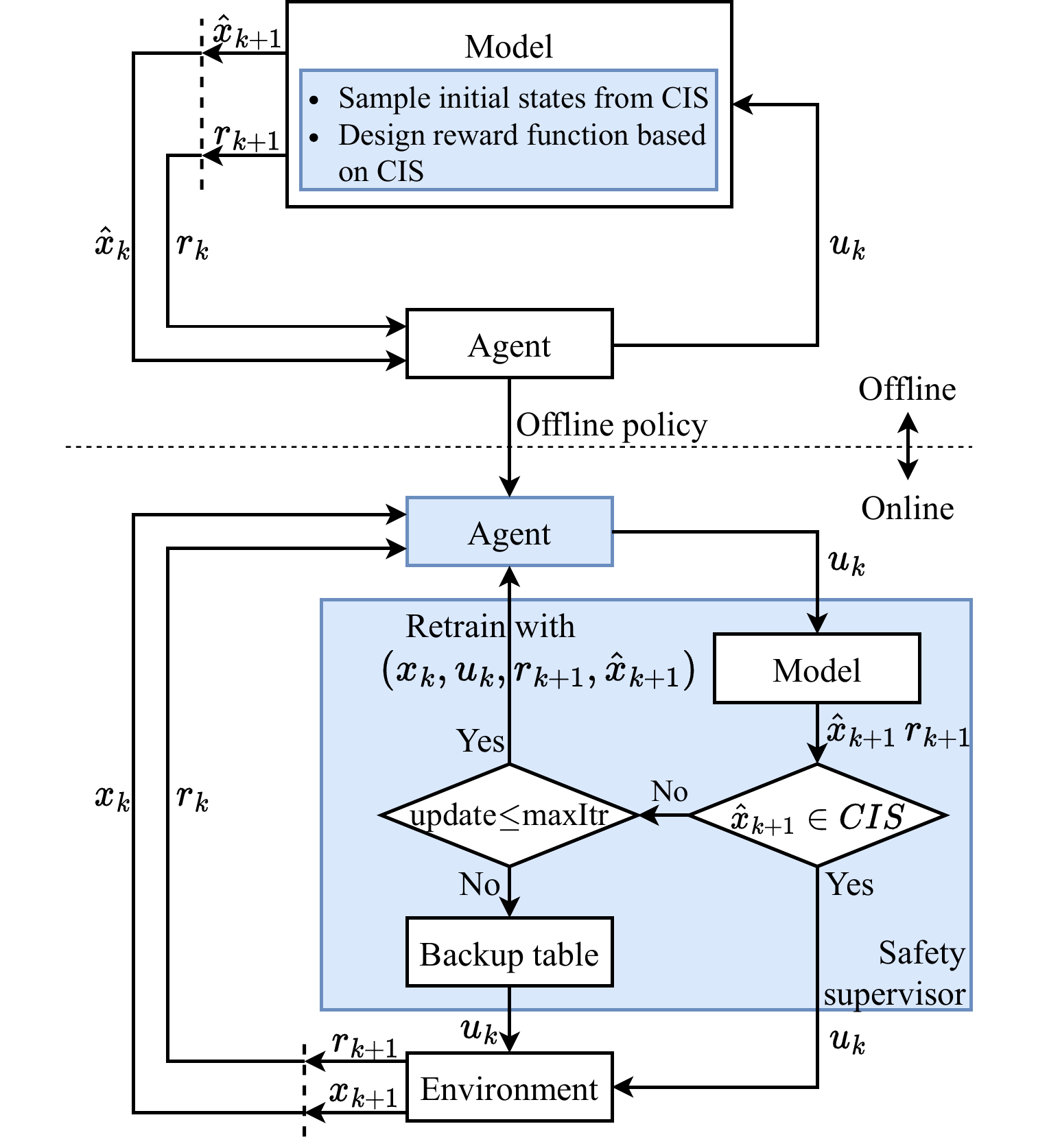}
		\caption{Flow diagram of the proposed RL training approach.}
		\label{fig:proposed_alg} 
	\end{figure}
	
	\subsection{Offline training} \label{sec:proposed_offline}
	
	It is assumed that a CIS of system~\eqref{eqn:nonlinear}, $R$, is available before the training of the RL. It is preferred that the CIS is the maximum CIS of~\eqref{eqn:nonlinear} within the physical constraint $X$, which ensures that the RL can explore within the biggest feasible and stable operating region. Meanwhile, we note that the maximum CIS is not a requirement in the proposed approach, any CIS can be used in the proposed approach. 
	
	In the proposed RL offline training, the system model~\eqref{eqn:nonlinear} is used in the training of the RL with $w=\textbf{0}$. In this section, we focus on demonstrating the concept of the proposed approach and assume that there is no model uncertainty. The CIS information is used in two ways. First, the CIS is used to penalize the RL agent when it drives the system state outside of the CIS. This can be achieved by designing the reward function appropriately. A discrete reward function shown in equation~\eqref{eqn:reward_general} may be used. 
	\begin{equation} \label{eqn:reward_general}
		r({x,u})= 
		\begin{cases}
			r_1& \text{if } {\hat{x}_{k+1} \in R}\\
			r_2              & \text{otherwise}
		\end{cases}
	\end{equation}
	where $R\subseteq X$ denotes the CIS of system~\eqref{eqn:nonlinear} used in the RL training, $r_1 \in \mathbb{R}$ and $r_2 \in \mathbb{R}$ denote the reward values associated with the prescribed action $u_k$ based on the current state {$\hat{x}_k$}. Note that $r_1$ should be greater than $r_2$ ($r_1 > r_2$) to guide the RL to prescribe control actions that can maintain the system state within the CIS. Specifically, at the time instant $k$, the system is at state {$\hat{x}_k$}. When the RL prescribes the control action $u_k$, it is sent to the model and the model will propagate into the next time instant and obtain {$\hat{x}_{k+1}$}. If {$\hat{x}_{k+1}$} is within the CIS (${\hat{x}_{k+1}}\in R${~by checking equation~\eqref{eqn:polytope}}), the RL will receive a higher reward $r_1$; if {$\hat{x}_{k+1}$} is outside of the CIS ({$\hat{x}_{k+1}\notin R$}), indicating the prescribed action resulting an unstable operation, the RL receives a relatively lower reward (or penalty) $r_2$. 
	
	Moreover, the CIS is used for initial state sampling in RL offline training. In RL offline training, an RL needs to sample the initial state of the system randomly for many times. Typically, the RL is restricted to sample the initial condition within the physical constraint set $X$. In the proposed approach, instead of using $X$, we propose to sample the initial state {$\hat{x}_0$} of the system within the CIS $R$. If the CIS is the maximum CIS, and if the system starts from an initial state outside of the CIS, the RL is not able to stabilize the system and drive the system back into the CIS. Such a case indeed does not provide too much information for learning a policy that ensures the stability. If the system starts within the CIS, then RL is able to find a control action to stabilize the system and learn the optimal and non-optimal actions based on the reward function~\eqref{eqn:reward_general}. Therefore, the sampling efficiency can be improved by restricting the RL to sample initial states within the CIS. This will be demonstrated in the simulations.
	
	Another technique we propose in the offline training is to reset the state to its previous value when the state is outside of the CIS. Assume that at time $k$, {$\hat{x}_k\in R$}. The RL agent prescribes a control action $u_k$ and drives the system state to be outside of the CIS; that is, {$\hat{x}_{k+1}\notin R$}. In such a case, the RL will get a lower reward $r_2$ according to~\eqref{eqn:reward_general}. Since once the system state is outside of the CIS, there is no control action that can drive the system back to the CIS (if the CIS is the maximum one within $X$) and the system becomes unstable, the interaction between the RL and the system will not further bring much useful information towards learning the optimal control law. Therefore, we propose to reset the state to its previous value; that is, set {$\hat{x}_{k+1}=\hat{x}_k$} and then continue the training process. By implementing this state reset technique, the RL learns from this failure experience and will get second or more chances to learn at the same state {${\hat{x}}_k$} towards the stable and optimal policy. {Upon initial inspection, the state reset technique may seem unconventional. However, it shares a similar concept with the gridworld example used often in RL~\cite{sutton2018reinforcement}.}
	
	\begin{figure}
		\centering
		\includegraphics[width=0.3\textwidth]{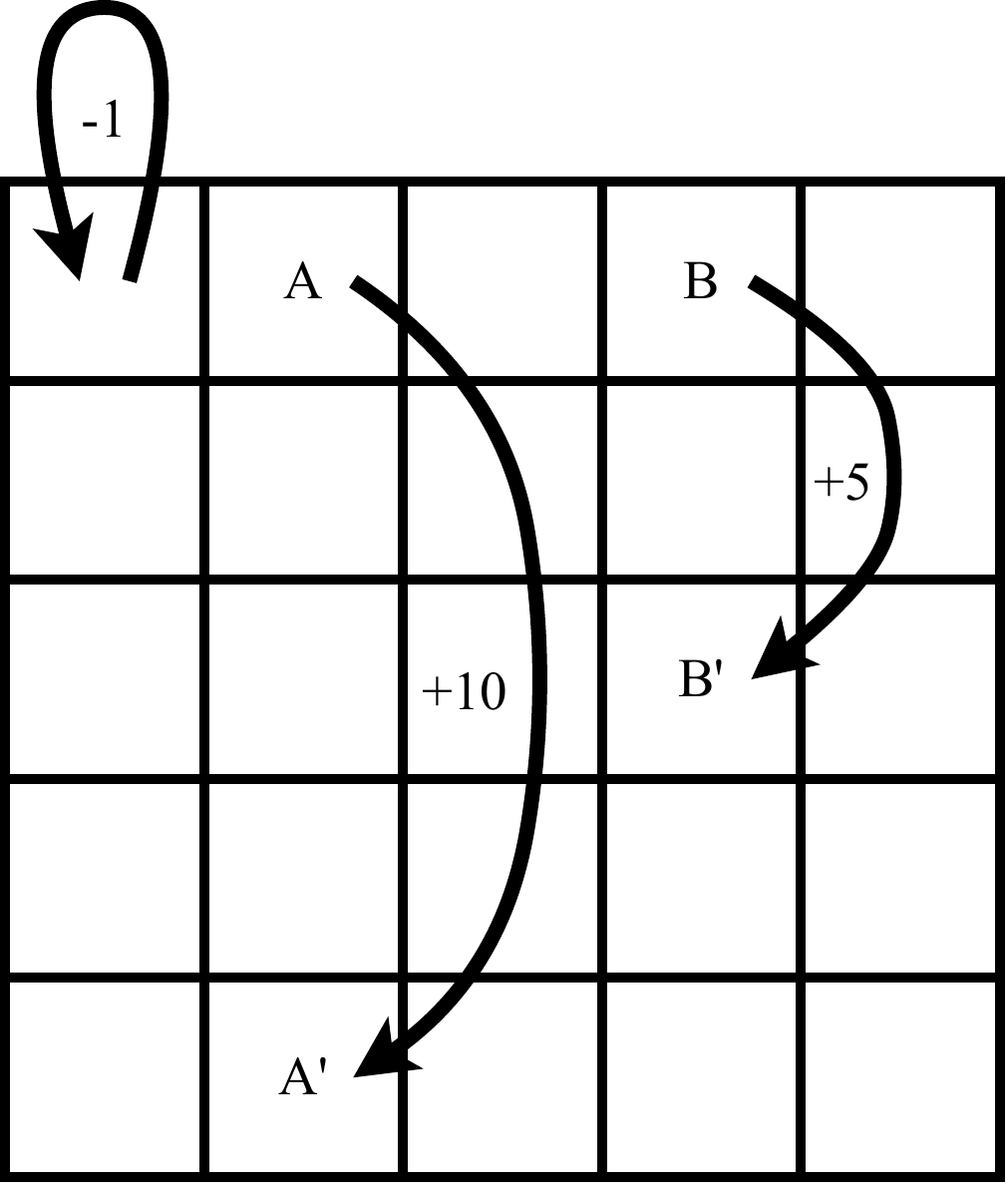}
		\caption{Gridworld.}
		\label{fig:grid_world} 
	\end{figure}
	{
		Figure~\ref{fig:grid_world} shows one gridworld example, in which the agent navigates a $5 \times 5$ gridworld with four possible actions: north, south, east, and west. By default, taking an action yields a reward of $0$. However, when the agent is located at either state $A$ or $B$, any action taken transitions it to $A'$ or $B'$ and earns a reward of $+10$ and $+5$, respectively. The common ground between this example and the state reset technique lies in their treatment of states near the boundary. In the gridworld example, as stated in \cite{sutton2018reinforcement}, actions that would move the agent off the grid result in its location remaining unchanged, accompanied by a reward of $-1$. This aligns precisely with the design of the state reset technique, because in both cases, when the prescribed action $u_k$ is unsafe, the successive sampled data, represented by equation~\eqref{eqn:data_tuple}, are all starting with the same values of $x_k$. The only distinction is that the boundaries of the grid in the example are clearly defined, while such explicit boundaries generally do not exist in real-world applications. In this work, the CIS effectively serves as the boundary, thus addressing this challenge in real-world scenarios.}
	
	{In summary, the incorporation of CIS facilitates the reward function design (affecting $r_{k+1}$ in data tuple~\eqref{eqn:data_tuple}), the initial state sampling (affecting $x_{k}$ in~\eqref{eqn:data_tuple} with $k=0$), and the state reset technique (influencing $x_k$ in~\eqref{eqn:data_tuple}). It is worth mentioning that these strategies do not modify RL algorithms themselves, such as updating the value function and policy network, etc. Instead, they solely provide the guidance on the sampling of data tuples~\eqref{eqn:data_tuple}. Hence, the proposed algorithm is applicable to a wide range of RL algorithms and the CIS guides the data sampling and is in favor for guaranteeing the stability.}
	
	\subsection{Online training and stability guaranteed implementation strategy} \label{sec:proposed_stability}
	
	After offline training, the RL learns a policy and the RL policy is saved as a pre-calculated controller for online implementation. However, due to the sampling nature of the RL training, it is impossible for the offline trained RL agent to encounter all situations. Therefore, the offline learned policy may not guarantee the closed-loop stability. To address this issue, we propose to implement the RL policy using a stability guaranteed strategy and to further train the RL policy online when a new situation is encountered. As shown in Figure~\ref{fig:proposed_alg}, a Safety Supervisor is placed in between the RL agent and the environment. The detailed description of the Safety Supervisor is shown below.
	
	Let us consider the current sampling time is $k$. With the state feedback ${x}_k$, the RL prescribes the control action $u_k$ according to the learned policy. Based on $u_k$ and the system model~\eqref{eqn:nonlinear}, the state {$\hat{x}_{k+1}$} at the next sampling time is predicted. If the predicted state {$\hat{x}_{k+1}$} is within the CIS ({$\hat{x}_{k+1}\in R$}), then the control action $u_k$ is actually applied to the system; if the predicted state {$\hat{x}_{k+1}$} is outside of the CIS ({$\hat{x}_{k+1}\notin R$}), then the RL is switched again to the training mode{, the reward $r_{k+1}$ is calculated following equation~\eqref{eqn:reward_general}} and the policy is updated with the new interaction experience $(x_k,u_k,r_{k+1},{\hat{x}_{k+1}})$. The updated policy will prescribe the updated action based on the state $x_k$ again {following the state reset technique shown in the previous section}. Unless the new policy guarantees that the predicted state {$\hat{x}_{k+1}$} is within the CIS, the agent will keep learning at the current state $x_{k}$ until a pre-determined maximum number of iterations (\text{$maxItr$}) is reached. {It is worth mentioning that the retraining technique keeps the RL algorithms intact as what offline training does. In the same way, the CIS provides guidance for sampling data tuples~\eqref{eqn:data_tuple}}. This online training/updating approach can significantly enhance the safety of the RL. However, the closed-loop stability is still not guaranteed. It is possible that within the maximum iteration $maxItr$, the online updating of the RL may not converge to a stable action (the RL cannot find a stable action for the particular state $x_k$ within $maxItr$ online updates). It should be pointed out that when $maxItr$ is very large, the online training is expected to find a safe action for every state it encounters since the CIS provides the guarantee of the existence of a safe action for all the states within it. {The stability proof will be shown later.} 
	
	To address the above issue and to guarantee the closed-loop stability, we propose the use of a backup table to save the safe actions for the states such that the above online training fail to find a safe action within $maxItr$ iterations. The safe actions can either be found using another stabilizing but not optimal control law, or by sampling the control action space randomly, or by leveraging the information contained in the explicit form of the CIS. Note that in some CIS construction algorithms \cite{decardi2022graph}, the corresponding safe action space for each state is also found, which can be taken advantage of in creating the backup table. This approach provides a safety guarantee for the RL.

	\begin{algorithm}
		\KwIn{$x_k$, $k$, $maxItr$}
		\KwOut{Safe $u_k$}
		$notSafe \gets True$, $update=1$\;
		\While{notSafe}{
			Calculate $u_k$ at $x_k$ based on the learned RL policy\;
			Based on the model and $u_k$, predict {$\hat{x}_{k+1}$ and $r_{k+1}$}\;
			\eIf{${\hat{x}_{k+1}} \in R$}{$notSafe \gets False$\;}
			{	
				\eIf{$update \leq maxItr$}{Update RL policy with $(x_k,u_k,r_{k+1},{\hat{x}_{k+1}})$\;
					$update \leftarrow update+1$\;}
				{Get safe action $u_k$ from the backup table\; $notSafe \gets False$\;}		
			}
		}
		Apply $u_k$ to system~\eqref{eqn:nonlinear} and obtain ${x}_{k+1}$\;
		Reinitialize the algorithm with $k\leftarrow k+1$\vspace{2mm}
            \caption{Safety Supervisor in online implementation for stability guarantee}\label{alg:two}
	\end{algorithm}
	
	The proposed stability guaranteed implementation strategy and online training is summarized in Algorithm~\ref{alg:two}. In the algorithm, the parameter $maxItr$ can be defined by the user to balance between the computational complexity and optimality of the RL agent. A larger value will allow the agent to be trained on a state for a longer time, which can potentially result in a better safety performance. However, this comes at the cost of increased online computational complexity. On the other hand, a small value, or even zero, will ensure stability and online implementation feasibility given that the backup table is designed well. However, it may not achieve optimal performance when relying on the backup plan for safe actions, because the selection among safe actions at the current state does not consider the optimality. One more factor to consider in picking, $maxItr$, is the sampling time of the system (time interval between two discrete time instants). It should be ensured that within one sampling time a stabilizing control action can be prescribed which indeed limits the maximum value of $maxItr$. 
	
	Note that in the above discussion, the primary focus was on maintaining the system state within the CIS ($x_k\in R$), which is also the concept of stability considered in this work. Since $R \subseteq X$, it also ensures that the state constraint ($x_k\in X$) is satisfied. One interesting feature of the proposed approach is that maintaining $x_k\in R$ is handled through the incorporation of the CIS in the offline training and the online implementation. This provides flexibility in the proposed approach to incorporate other control objectives such as set-point tracking, zone tracking or economic optimization in the design of the reward function $r$.
	
	\subsection{Control objectives}\label{sec:proposed_control_objectives}
	This section presents how the proposed algorithm integrates three types of control objectives, including set-point tracking, economic optimization, and zone tracking. As simple as distinguishing two constants ($r_1$ and $r_2$) in~\eqref{eqn:reward_general} to enable RL understand the CIS boundary, the control objectives can be taken into account straightforwardly by replacing the constant $r_1$ with the corresponding cost functions. Hence, while maintaining the same structure as equation~\eqref{eqn:reward_general}, the modification on $r_1$ is documented as follows:
	
	\textbf{Set-point tracking}
	\begin{equation}
		r_1 = -\norm{x_k-x_s}_p^q
	\end{equation}
	where $x_s$ represents the set-point that needs to be tracked, while the subscript $p$ and the superscript $q$ denote the type of $l_p$ norm and the exponentiation design, respectively. A negative sign is added to the equation because the problem is formulated as a maximization problem in the context of RL.
	
	\textbf{Economic maximization}
	\begin{equation}
		r_1 = l_e(x, u)
	\end{equation}
	where the function $l_e: \mathbb{R}^n\times \mathbb{R}^m \rightarrow \mathbb{R}$ is used to evaluate the economic performance of the process based on the current state $x$ and input $u$. A higher value of $l_e$ indicates better performance. In addition, there is no linearity restriction on $l_e$.
	
	\textbf{Zone tracking}
	\begin{equation}
		\begin{aligned}
			r_1 = -\min_{x_z} \quad & \norm{x_k-x_z}_p^q \\
			\textrm{s.t.} \quad & x_z \in X_z
		\end{aligned}
	\end{equation}
	where the variable $x_z$ is a slack variable and $X_z \subseteq R$ represents the target zone. When the state $x_k$ is within the target zone $X_z$, the reward function $r_1$ achieves at its highest value of 0 with $x_z=x_k$. On the other hand if $x_k$ is not in $X_z$, the distance between $x_k$ and $X_z$ is computed and used to penalize the RL algorithm. The parameters $p$, $q$ and the negative sign serve the same purpose as in the set-point tracking.
	
	It is important to note in order to achieve stability while fulfilling the control objectives, the minimum value of $r_1$ must be greater than $r_2$ ($r_1>r_2$). In addition, the equations presented above serve as an illustration of how different control objectives can be incorporated into the proposed CIS-enhanced RL framework.  However, they are not intended as the best or final version of cost design, and users are encouraged to tailor them to the specific nature of their own projects.
	
	\subsection{{Stability proof}}
	The main objective of this research is to develop a CIS-enhanced RL approach that ensures closed-loop stability during real-time implementation, enabling its application in safety-critical processes. In this section, we provide a formal proof of stability for the proposed online implementation under deterministic conditions.
	\begin{theorem}\label{the:1}
		Consider a system described by equation~\eqref{eqn:nonlinear} with $W=\textrm{\O}$ and a known CIS $R$. Let the initial state $x_{0}\in R$. Let an RL agent be trained as discussed in Section\ref{sec:proposed_offline}, the exploration rate of the RL be non-zero, and a Safety Supervisor be designed according to Algorithm~\ref{alg:two} with $maxItr=\infty$. Then, the proposed online implementation of the RL agent can guarantee that system~\eqref{eqn:nonlinear} is maintained within the CIS $R$ all the time; that is, $x_k\in R,~\forall k\geq 0$ .
	\end{theorem}
	\textbf{Proof}:
	Consider the state at time $k$, $x_k$, and assume that $x_k\in R$. Two cases are considered depending on the safety of the RL agent's action $u_k$:
	\begin{enumerate}
		\item Safe $u_k$: According to Definition~\ref{def:cis}, for any $x_k\in R$, a safe action $u_k$ exists that ensures $x_{k+1}\in R$. In other words, if $x_k\in R$ and action $u_k$ is safe, then $x_{k+1}\in R$.
		\item Unsafe $u_k$:  If the RL agent prescribes an unsafe action $u_k$, according to the online implementation, the RL will be retrained online. Since $x_k\in R$, there always exists a safe action based on Definition~\ref{def:cis}. If $maxItr=\infty$, meaning that the RL agent is allowed to iterate infinitely at time instant $k$, the state reset technique enables the RL agent to collect data $(x_k,u_k,r_{k+1},\hat{x}_{k+1})$ at failed state $x_k$ an infinite number of times. By utilizing the reward function~\eqref{eqn:reward_general}, it detects violations of $R$ and reflects the violation on $r_{k+1}$ in the sampled data. If the exploration rate remains non-zero, the RL agent exhaustively explores all possible actions, allowing the discovery of a safe action. Consequently, $x_{k+1}\in R$ can be achieved.
	\end{enumerate}
	The above implies that if $x_k\in R$, the proposed online implementation of the RL can ensure that $x_{k+1}\in R$. If the initial condition of the system $x_0 \in R$, applying the above conclusion recursively, it can be proved that $x_k\in R$ for all $k\geq 0$. This proves Theorem~\ref{the:1}. $\square$

\section{Proposed approach - stochastic case}\label{sec:proposed_stoc}
In the presence of model uncertainty, it is possible that some states within the CIS are no longer control invariant, meaning that there is no action that can maintain the next state within the CIS. To account for this, the proposed design can be adapted to consider an RCIS, which excludes the states, becoming control variant due to uncertainties, from the safe zone for RL exploration. 
The framework for the stochastic case remains the same as for the deterministic case, with essential modifications required for offline learning and online implementation documented below.

\subsection{Offline training}\label{sec:proposed_stoc_offline}
The offline training only requires replacing the CIS with the RCIS, while keeping the same initial state sampling, reward function design, and state reset technique. Hence the reward function~\eqref{eqn:reward_general} is modified to the following with $r_1>r_2$:
\begin{equation} \label{eqn:reward_general_stoc}
	r({x,u})= 
	\begin{cases}
		r_1& \text{if } {\hat{x}_{k+1} \in R_r}\\
		r_2              & \text{otherwise}
	\end{cases}
\end{equation}
Under model uncertainty, states within the CIS $R$ may not be control invariant because this property is only preserved in the deterministic case. If a state that is not control invariant is sampled as an initial state or the RL prescribes an action to visit the such state, the RL agent will receive incorrect reward feedback if $R$ is utilized. By using the $R_r$ that contains only control invariant states under the stochastic case, these incorrect states can be excluded and the agent can receive the correct penalty when visiting them.

\subsection{Online training and stability guaranteed implementation strategy}\label{sec:proposed_stoc_online}
In the online implementation, the main difference between the stochastic and deterministic cases is that the predicted state $\hat{x}_{k+1}$ by the model may differ from the actual response $x_{k+1}$ of the environment. In the deterministic case, the safety of the action $u_k$ is checked by verifying whether the predicted state $\hat{x}_{k+1}$ is within the safe region $R$. If $\hat{x}_{k+1} \in R$, then it can be inferred that $x_{k+1} \in R$ and $u_k$ is safe. However, in the stochastic case, an action $u_k$ leading to a predicted state $\hat{x}_{k+1}\in R_r$ does not guarantee that $x_{k+1}\in R_r$, since we cannot predict the value of the uncertainty. Therefore, a more rigorous approach is required to ensure the safety of the action and cannot solely rely on the predicted state $\hat{x}_{k+1}$.

The proposed modification to the Model block in the Safety Supervisor of Figure~\ref{fig:proposed_alg} involves an optimization problem that promotes the worst disturbance $w_k$ that, in combination with the action $u_k$, can drive the model $\hat{x}_{k+1}$ outside of $R_r$. 
By considering the worst disturbance, the optimization problem provides a more robust and conservative approach to ensure that the action $u_k$ is safe and that the resulting state $\hat{x}_{k+1}$ is guaranteed to remain within $R_r$, inferring $x_{k+1}\in R_r$.
If, for all $w_k$ the optimization problem visited, $\hat{x}_{k+1} \in R_r$, then the action $u_k$ is deemed safe and $x_{k+1}$ is guaranteed to remain within $R_r$.

The optimization problem is formulated to maximize the probability that the predicted state $\hat{x}_{k+1}$ is outside of $R_r$ and shown below:
\begin{equation}\label{eqn:opt_concept}
	\begin{aligned}
		J^* = \max_{w_k} \quad & P(\hat{x}_{k+1} \notin R_r)\\
		\textrm{s.t.} \quad & \hat{x}_{k+1} = f(x_k, u_k, w_k) \\
		\quad & w_k \in W
	\end{aligned}
\end{equation}
where the optimization variable $w_k$ is bounded by the known set $W$. The objective function $P(\hat{x}_{k+1} \notin R_r)$ denotes the probability of $\hat{x}_{k+1}$ being outside of $R_r$. If the optimal value $J^*=1$, then there exists a disturbance $w_k$ that can be applied to the system to make the actual state $x_{k+1}$ violate the safe set $R_r$. The probability serves as a conceptual objective, meaning that it does not necessarily need to be explicitly defined or calculated. Instead, in this work, the explicit form of the RCIS is used to define the probability indirectly. Users can define their own probability objective function, either directly or indirectly, depending on the nature of the safety constraints and the specific requirements of the application.

{In this work, substituting} the constraint in equation~\eqref{eqn:polytope} into equation~\eqref{eqn:opt_concept}, the optimization problem is shown below:

\begin{equation}\label{eqn:opt_max}
	\begin{aligned}
		J^*=\max_{w_k} \quad & \max(A\hat{x}_{k+1}-b)\\
		\textrm{s.t.} \quad & \hat{x}_{k+1} = f(x_k, u_k, w_k) \\
		\quad & w_k \in W
	\end{aligned}
\end{equation}
where the $\max(\cdot)$ operator in the objective function is used to convert the $A\hat{x}_{k+1}-b$ from a vector to a scalar and, more importantly, to select the element with the maximum value, which indicates the boundary with the highest possibility of being violated. The maximum optimization promotes the violation by trying to maximize the largest element in $A\hat{x}_{k+1}-b$ through modifying the decision variable $w_k$. If the optimal value $J^*>0$, it implies that the prescribed action $u_k$ could result in $\hat{x}_{k+1}$ being outside of the $R_r$. In this case, $u_k$ should not be passed to the environment and the RL agent should be penalized and retrained online. 
%{\color{red}Similarity to Hamilton-Jacobi Reachability Analysis}

The optimization problem~\eqref{eqn:opt_max} is non-smooth due to the presence of the $\max$ operator, which makes it unsuitable for gradient-based optimization algorithms that require differentiable objective functions. To overcome this issue, a new variable $z$ is introduced to represent the maximum element in $A\hat{x}_{k+1}-b$, leading to a reformulated optimization problem as follows:
\begin{align}\label{eqn:opt_milp}
	J^*=\max_{z, y_i, w_k} \quad & z\\
	\textrm{s.t.} \quad & z \leq A_i\hat{x}_{k+1}-b_i+My_i,~i=1,...,c \label{eqn:opt_milp_1}\\ 
	\quad & \sum_{i=1}^{c}y_i = c-1 \label{eqn:opt_milp_2}\\
	\quad & y_i = \{0, 1\},~i=1,...,c\\
	\quad & \hat{x}_{k+1} = f(x_k, u_k, w_k) \label{eqn:opt_milp_4}\\
	\quad & w_k \in W \label{eqn:opt_milp_5}
\end{align}
where there are $c$ binary variables $y_i$ corresponding to the $c$ affine constraints in $A\hat{x}_{k+1}-b$, with each $y_i$ indicating whether the $i$-th constraint in~\eqref{eqn:opt_milp_1} is active ($y_i=0$) or inactive ($y_i=1$). Constraint~\eqref{eqn:opt_milp_2} forces exactly $c-1$ constraints to be inactive, with the remaining one being active. The active constraint with $y_i=0$ results in its corresponding equation in~\eqref{eqn:opt_milp_1} converted to $z\leq A_i\hat{x}_{k+1}-b_i$, which provides an upper bound on the maximum element in $A\hat{x}_{k+1}-b$. Since the optimal value $J^*$ represents the maximum element in $A\hat{x}_{k+1}-b$, the constraint $J^*\leq A_i\hat{x}_{k+1}-b_i$ is no longer satisfied for the inactive constraints. To ensure that $J^*\leq A_i\hat{x}_{k+1}-b_i+My_i$ holds for all constraints in~\eqref{eqn:opt_milp_1}, an auxiliary positive parameter $M\in \mathbb{R}{+}$ is introduced and set sufficiently large. {It is important to note that the optimization process can also be applied during offline training by sacrificing time efficiency in order to obtain an improved offline agent that can potentially save time during online implementation. Hence, the decision of whether to include the optimization process in the offline training depends on the specific time constraints of the scenario. In the forthcoming results, we will demonstrate that even if the optimization process is not incorporated during offline training, the online implementation still ensures stability. Additionally, we will provide a detailed study on the time usage of the implementation.}

The optimization problem~\eqref{eqn:opt_milp} is classified as a mixed-integer problem (MIP) and can be further categorized as a mixed-integer linear problem (MILP) based on the structure of equation~\eqref{eqn:opt_milp_4}, where $w_k$ only involves linear operations. It is worth noting that nonlinear operations between $x_k$, $u_k$, and $w_k$, such as $x_k/w_k$ or $u_k^2w_k$, etc., do not affect the MILP, as $x_k$ and $u_k$ are constant in the optimization. The advantage of MILP is that there exist mature and computationally efficient solvers \cite{sioshansi2017optimization}, which means that implementing the proposed method online will not introduce significant computational complexity issues.

{After solving the MILP problem, if $J^*\leq 0$, then $u_k$ is safe and can be passed to the Environment. Otherwise, the optimal solution $w_k^*$, i.e., the worst disturbance, is used to predict $\hat{x}_{k+1}$ in equation~\eqref{eqn:opt_milp_4} in data tuple collection for retraining purpose.}

\subsection{{Stability proof}}
In this section, we present a formal proof of stability for the proposed online implementation with the presence of uncertainty.
\begin{theorem}\label{the:2}
Consider a system described by equation~\eqref{eqn:nonlinear} with a bounded uncertainty $W$ and a known RCIS denoted as $R_r$. Let the initial state $x_{0}\in R_r$. Let an RL agent be trained as discussed in Section~\ref{sec:proposed_stoc_offline}, the exploration rate of the RL be non-zero, and a Safety Supervisor be designed according to Algorithm~\ref{alg:two} with the addition of the MILP formulation~\eqref{eqn:opt_milp}-\eqref{eqn:opt_milp_5} and with $maxItr=\infty$. Then, the proposed online implementation of the RL agent can guarantee that system~\eqref{eqn:nonlinear} is maintained within the RCIS $R_r$ all the time; that is, $x_k\in R_r,~\forall k\geq 0$.
\end{theorem}
\textbf{Proof}: Consider the state at time $k$, $x_k$, and assume that $x_k\in R_r$. We consider two cases depending on the safety of the RL agent's action $u_k$:
\begin{enumerate}
	\item Safe $u_k$: According to Definition~\ref{def:rcis}, for any $x_k\in R_r$, there exists a safe action $u_k$ that ensures $x_{k+1}\in R_r$. In other words, if the state $x_k\in R_r$ and the action $u_k$ is safe, then $x_{k+1}\in R_r$.
	\item Unsafe $u_k$: If the RL agent prescribes an unsafe action $u_k$, according to the online implementation, the RL will be retrained online. Since $x_k\in R_r$, there always exists a safe action based on Definition~\ref{def:rcis}. If the RL agent is allowed to iterate infinitely at time instant $k$ ($maxItr=\infty$), the state reset technique enables the RL agent to collect data $(x_k,u_k,r_{k+1},\hat{x}_{k+1})$ at the failed state $x_k$ an infinite number of times. By utilizing the reward function~\eqref{eqn:reward_general_stoc} with the MILP formulation~\eqref{eqn:opt_milp}-\eqref{eqn:opt_milp_5}, violations of $R_r$ are detected and reflected in the reward signal $r_{k+1}$. The MILP problem ensures the accuracy of safety of $u_k$ with the presence of uncertainty and provides the correct calculation of $r_{k+1}$. If the exploration rate remains non-zero, the RL agent exhaustively explores all possible actions, enabling the discovery of a safe action. Consequently, $x_{k+1} \in R_r$ can be achieved.
\end{enumerate}
The above implies that if $x_k \in R_r$, the proposed online implementation of the RL can ensures that $x_{k+1}\in R_r$. If the initial condition of the system $x_0\in R_r$, applying the above conclusion recursively, it can be proved that $x_k\in R_r,~\forall k\geq 0$. This proves Theorem~\ref{the:2}. $\square$

\section{Simulation results and discussion - without uncertainty}

\subsection{Process description}

In order to study the sampling efficiency and the closed-loop stability guarantee of the proposed RL training and implementation approach, the application of the proposed approach to the control of a continuously stirred tank reactor (CSTR) is considered in this section. The reaction happening inside of the reactor is an irreversible and exothermic reaction with first-order reaction rate. The CSTR is also installed with a cooling jacket outside of it for maintaining the temperature of the reaction mixture. The mathematical model {contains} two nonlinear ordinary differential equations with the following representation {\cite{decardi2022robust}}:
\begin{eqnarray*}\label{eqn:cstr_model}
	&& \frac{dc_A}{dt}=\frac{q}{V}(c_{Af}-c_A)-k_{0}exp(-\frac{E}{RT})c_{A} \\
	&&   \frac{dT}{dt} = \frac{q}{V}(T_f-T)+\frac{-\Delta H}{\rho c_p}k_{0}exp(-\frac{E}{RT})c_{A}+\frac{UA}{V\rho c_{p}}(T_c-T)
\end{eqnarray*}
where $c_A$ $(mol/L)$ and $T$ $(K)$ denote the concentration of the reactant and the temperature inside of the reaction mixture, respectively. $c_{Af}$ $(mol/L)$ and $T_f$ $(L)$ represent the concentration of the reactant and temperature of the inlet stream. $T_c$ $(K)$ is the temperature of the coolant stream used for cooling the reactor temperature. The remaining parameters are summarized in {Table~\ref{tbl:parameters}}. The parameter $q$ is the inlet and outlet flow {rate} of the reactor, $V$ is the volume of the reaction mixture, $k_0$ is the pre-exponential factor of the Arrhenius rate constant, $E$ represents the activation energy required by the reaction, $R$ denotes the universal gas constant, $\Delta H$ is the change of the enthalpy used for approximating the change of internal energy of the reaction, $\rho$ is the density of the reaction mixture, $c_p$ denotes the specific heat capacity of the reaction mixture, $UA$ is the heat transfer coefficient between the reactor and the cooling jacket.

\begin{table}[t]
	\centering
	\renewcommand{\arraystretch}{1.5}
	\footnotesize
	\caption{Parameters of the CSTR model.} 
	\begin{tabular}{ccc} 
		\hline
		Parameter &       Unit &      Value \\
		\hline
		$q$ &      $L/min$ &        $100$ \\
		
		$V$ &          $L$ &        $100$ \\
		
		$k_0$ &      $min^{-1}$ &        $7.2 \times 10^{10}$    \\
		
		$E/R$ &      $K$      &        $8750.0$    \\
		
		$- \Delta H$ &      $J/mol$ &   $5.0 \times 10^{4}$ \\
		
		$\rho$ &        $g/L$ &       $1000.0$ \\
		
		$c_p$ &       $J/gK$ &      $0.239$ \\
		
		$UA$ &     $J/minK$       &    $5.0 \times 10^4$        \\
		
		$c_{Af}$   & $mol/L$ & $1$\\
		$T_{f}$    & $K$ & $350$\\
		\hline
	\end{tabular}  \label{tbl:parameters}
\end{table}

In the following closed-loop control problem study, $c_A$ and $T$ are the two states of the system and $T_c$ is the manipulated variable. They are subject to the following physical constraints: 
\begin{eqnarray}
	0.0\leq c_A \leq 1.0 \label{cstrcon1}\\
	345.0\leq T \leq355.0 \label{cstrcon2}\\
	285.0 \leq T_c \leq 315.0
\end{eqnarray}

The control objective is to train an RL policy to maintain a stable operation of the CSTR such that the two states are maintained within the physical constraints shown in~\eqref{cstrcon1} and~\eqref{cstrcon2} all the time.

The maximum CIS of the CSTR was calculated using the graph-based algorithm developed in \cite{decardi2021computing}. The physical constraints and the calculated maximum CIS over the state space are shown in Figure~\ref{fig:representation_of_cis}. According to Figure~\ref{fig:representation_of_cis}, the CIS spans over the entire temperature space and shrinks over the concentration space. Hence, if the concentration of the reactant is lower than the minimum value of $c_A$ in CIS (the top left point of CIS), no matter how $T_c$ is manipulated by the controller, the system becomes unstable. The same observation applies to when the system has a concentration that is higher than the maximum value of $c_A$ in CIS (bottom {right} point of CIS). Since the calculated CIS is the maximum one, when the system is outside of the CIS, there is no feedback control law that is able to bring the system back into CIS again and the system state will eventually diverge. This implies that the physical constraints will be violated. 

\begin{figure}
	\centering
	\includegraphics[width=0.6\textwidth]{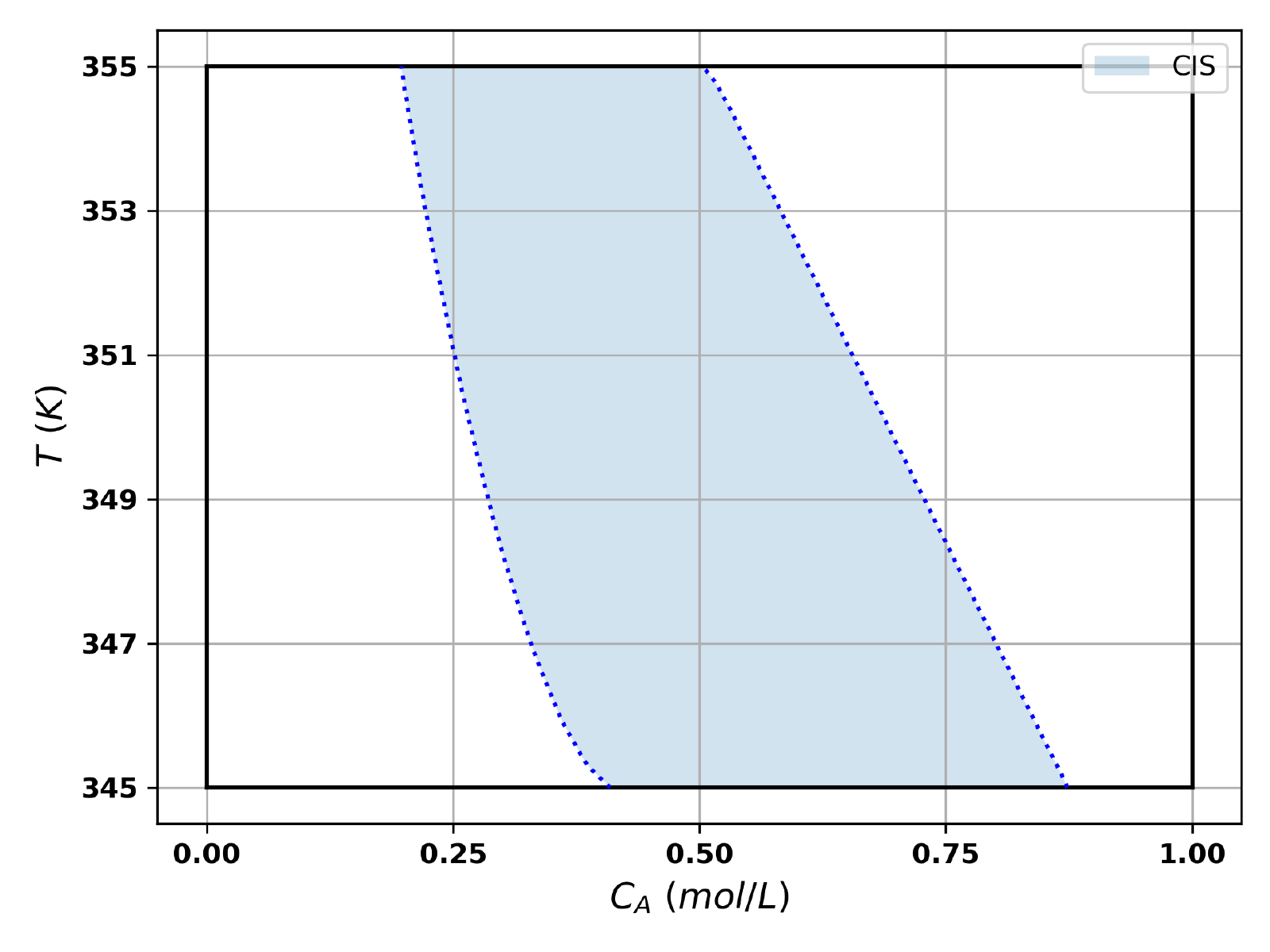}
	\caption{The physical constraints and the maximum CIS of the CSTR.}
	\label{fig:representation_of_cis} 
\end{figure}

\subsection{RL {offline} training setup and results} \label{sec:rl_train_setup}

In the training of the proposed RL, the maximum CIS was used. The proximal policy optimization (PPO) \cite{schulman2017proximal} was used as the optimization algorithm in RL training. Even though, the nature of the CSTR system was a continuous process, the RL agent was trained under the episodic setting. The reason is that when the system reaches its steady state as time goes by, the collected data, the steady state transition, will not provide any new information to the agent. Though the exploration-exploitation embedded in PPO algorithm may prescribe actions enhancing the exploration, meaning at the given steady state PPO will not prescribe the corresponding steady input, the probability of the exploration becomes extremely low. Hence, in order to favor the exploration, once the agent interacting with the environment for a user-defined number of time steps, the episode was terminated and a new initial state was sampled. During the experiment, 10,000 episodes and 200 steps per episode {were} used to train the RL agent. The batch size was defined as 10 episodes, meaning that the RL agent would be updated only {when all} 10 episodes {were} finished and the RL would learn from the data of the 10 episodes at once. The learning rate {was} defined as $10^{-4}$, discounted factor {was} $0.99$. It is noticed that the offline training takes overall 2,000,000 steps which might be computational expensive. However, the trained policy can be implemented online as a pre-calculated function which will require less online computation resources.

The consideration of CIS in the proposed RL training setup was reflected in two steps, the sampling of the initial states and the design of reward function. According to Section~\ref{sec:proposed_offline}, since the largest CIS is known, the initial state has to be inside of CIS to ensure the following states get the chance to be inside of CIS. Hence, all 10,000 initial states for 10,000 episodes were sampled within the CIS. In addition, because it {was} undesired for the system to enter the space outside of CIS, a discrete reward function was proposed as the following:
\begin{equation} \label{eqn:reward}
	r({x,u})= 
	\begin{cases}
		10,000& \text{if } {\hat{x}_{k+1} \in R}\\
		-1,000              & \text{otherwise}
	\end{cases}
\end{equation}

Based on aforementioned RL training setup, 20 offline training were executed in parallel and the learning performance was calculated as the average of performance over 20 training. The average training reward plot, representing the learning performance, is shown in Figure~\ref{fig:train_reward}. The orange horizontal line represents the maximum score each episode can achieve if the RL agent maintain the system within the CIS for all 200 steps{; the maximum score for each episode is $200 \times 10,000 = 2\times 10^6$}. The mean curve was calculated based on all 20 training and blue shaded area shows the one standard deviation. {Please note, in order to smooth out the fluctuations of the scores among episodes, the running average, which recursively calculated the average of scores of past 100 episodes, was used to plot the figure.} As RL agent interacting with the environment for more episodes, the score of episode increases, meaning RL agent is able to learn from the training. From the beginning to around 2,000$^{th}$ episodes, the RL agents has a higher learning rate with a larger variance between 20 training. After that, the learning slows down and gradually reaches the plateau with a decreased variance.

\begin{figure}
	\centering
	\includegraphics[width=0.6\textwidth]{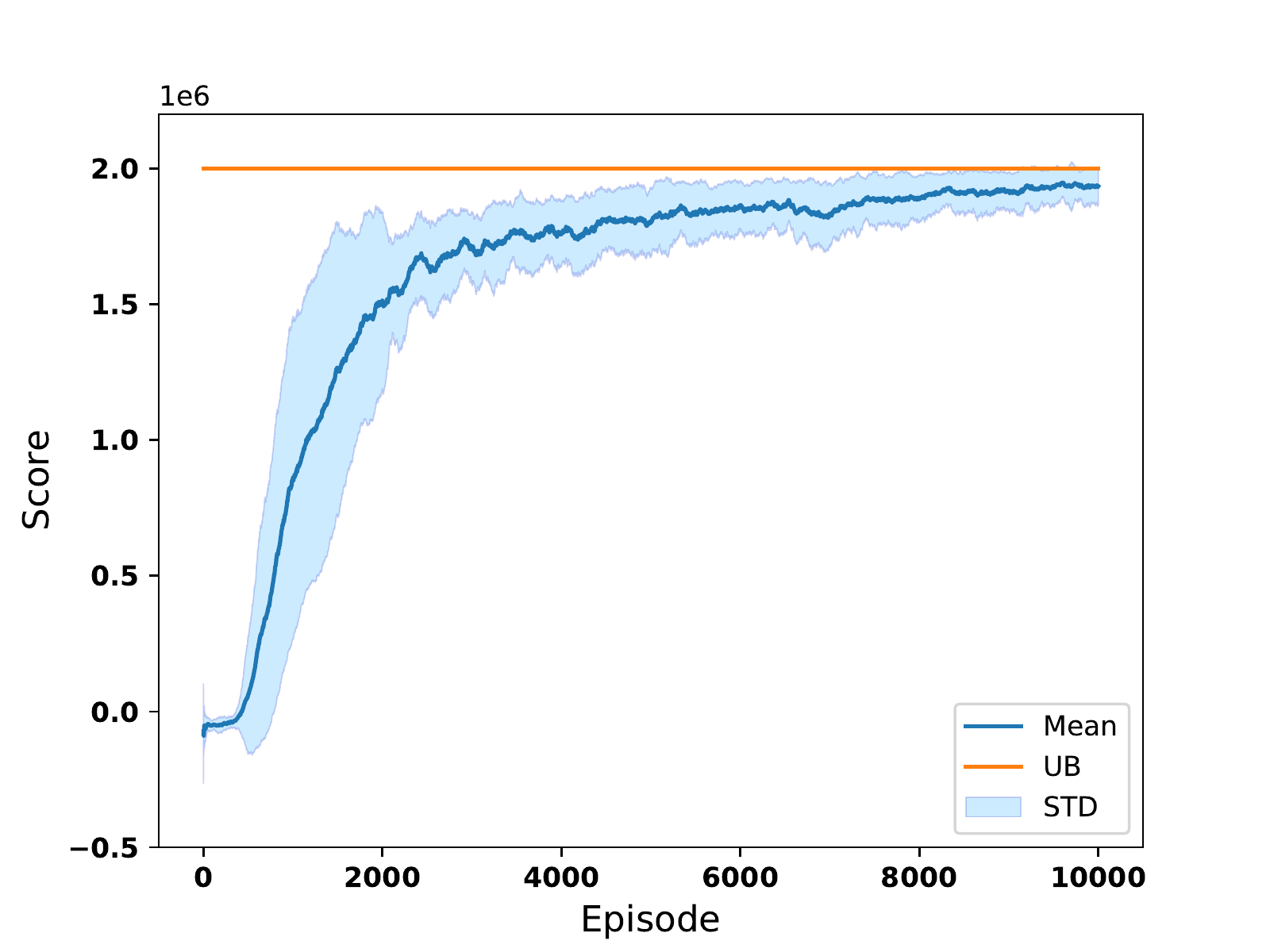}
	\caption{Average training score of proposed RL design}
	\label{fig:train_reward} 
\end{figure}

\subsection{RL testing setup and results} \label{sec:rl_test_setup}

Since in the RL training approach proposed in Section~\ref{sec:proposed_procedure}, an RL policy is trained offline before interacting with the environment online, it enables us to test the performance of the offline RL policy. The policy was tested with the model for 10,000 episodes and each episodes lasted for 200 steps. During the tests, the reward value was collected for evaluation purpose and not for learning purpose. Note that in all these tests, the control actions prescribed by the RL agents {were} applied directly to the CSTR and the proposed stability guaranteed online implementation strategy {was} not used. 

Table~\ref{tbl:test_fail_rate} summarizes the failure rates of the 20 trained RL policies. The failure is defined as when the policy cannot maintain the process within the CIS in an episode. According to the table, Run 18 is able to achieve a 0.02\% of failure rate. In other words, out of 10,000 episodes, there are only 2 episodes that the agent is failed to keep them inside of the CIS. Run 12 has the highest failure rate of 13.90\%. Overall, the proposed CIS enhanced offline RL training is able to achieve an average of 8.42\% failure rate. 

\begin{table}[t]
	\centering
	\renewcommand{\arraystretch}{1.5}
	\footnotesize
	\caption{Failure rates of RLs in testing mode.}
	\begin{tabular}{cccc} 
		\hline
		Run \# &      Failure rate (\%) & 	Run \# &      Failure rate (\%)\\
		\hline
		1 &      7.91 & 11&8.14\\  
		
		2 &     5.79 & 12&13.90\\
		
		3 &    3.41    & 13&13.88\\
		
		4 &     12.48     & 14&10.78\\
		
		5 &   2.25   & 15&12.92\\
		
		6 &  12.01     & 16&12.85\\  
		
		7 &    3.76   & 17&12.94\\
		
		8 &    4.21    & 18&0.02\\
		
		9 &     13.37     & 19&7.44\\
		
		10 &   4.39   & 20&6.04\\
		\hline
	\end{tabular}  \label{tbl:test_fail_rate}
\end{table}

% In terms of the initials states for 10,000 episodes, two groups of initial states were studied, the in-sample and the out-of-sample. The in-sample contained 10,000 initial states that were the same as those used for training the agent. The out-of-sample included initial states that were sampled differently from those used in training.

%\songrev{Figure~\ref{fig:test_reward} depicts the general testing performance over the 20 offline RLs. Along the x-axis, the blue curve does not have the rising trend as the one in Figure~\ref{fig:train_reward} because there is no learning process in the testing. For 10,000 different initial states, on average, the offline RL is able to achieve a score of 1.82$\times 10^6$, which is close to the maximum possible score. 
	
	%\begin{figure}[H]
	%	\centering
	%	\includegraphics[width=0.9\textwidth]{score_history_running_avg_test.png}
	%	\caption{Average testing score of proposed RL design}
	%	\label{fig:test_reward} 
	%\end{figure}
	
	\subsection{{Study of sampling efficiency}} \label{sec:sampling_efficiency}
	In order to study and quantify the benefits brought by utilizing the CIS in RL offline training, the sampling efficiency was studied. {The main objective is to assess the benefits derived from the explicit form of CIS.} The study was conducted by comparing the results shown in Section~\ref{sec:rl_train_setup} with an RL training setup that {did} not utilize the CIS information. Hence, in this RL without CIS training, 10,000 of initial states were sampled within the physical constraints and the reward condition was extended from CIS $R$ in equation~\eqref{eqn:reward} to the whole physical constraint $X$. In addition, the reset of the state was done when the system state was outside of the physical constraints. Other parameters remained the same. 
	
	Since two RL training setups had different reward functions, it was impossible to compare their training plots directly. {Hence, they were tested by comparing the failure rates introduced in Section~\ref{sec:rl_test_setup}}. The same 10,000 initial states were tested. Table~\ref{tbl:sampling_efficiency} shows that the RL with CIS is able to achieve 8.42\% of failure rate over 10,000 episodes and RL without CIS can only achieve 34.30\%. Therefore, using the same amount of data in training, utilizing the knowledge of CIS facilitates the learning process. Then both RL agents were trained with 20,000, 30,000, 40,000 and 50,000 episodes, then tested with the same 10,000 initial states that were used before. The failure rate of RL with CIS has a minor improvement to 4.84\% and that of RL without CIS has a relative bigger improvement to 14.58\%. However, the failure rate of RL without CIS trained using 50,000 episodes {(14.58\%)} is still higher than that of RL with CIS trained only using 10,000 episodes {(8.42\%)}. Hence, the utilization of CIS improves the sampling efficiency of RL training process.
	
	\begin{table}[t]
		\centering
		\renewcommand{\arraystretch}{1.5}
		\footnotesize
		\caption{Failure rates of RLs trained with different sizes of dataset.}
		\begin{tabular}{ccc} 
			\hline
			Training dataset size &      RL with CIS &      RL without CIS \\
			\hline
			10,000 &      8.42\% &     34.30\%    \\  
			
			20,000 &          8.30\% &        25.84\% \\
			
			30,000 &      6.55\% &        19.83\%    \\
			
			40,000 &       5.52\%     &        16.56\%    \\
			
			50,000 &    4.84\%   &   14.58\% \\
			\hline
		\end{tabular}  \label{tbl:sampling_efficiency}
	\end{table}
	
	\subsection{Study of {online} stability guarantee} \label{sec:stability_guarantee}
	
	{After assessing the offline training performance and the sampling efficiency}, one offline RL agent was saved and treated as the pre-calculated feedback controller. Then by following the algorithm proposed in Section~\ref{sec:proposed_stability}, the agent was implemented online. The agent interacted with the environment for 10,000 episodes and each episodes lasted for 200 steps. In order to examine the benefits of the proposed online implementation, RL Run 1 obtained from Section~\ref{sec:rl_train_setup} was picked, because a near average test performance {was observed}. 
	
	First of all, with the proposed online implementation strategy, the agent was able to maintain the system within the CIS for all 10,000 episodes. This is expected since the proposed online implementation is stability guaranteed. 
	
	Second, since the agent was retrained during its implementation, it was expected that the retrained RL agent would be able to achieve a better performance in terms of stability. Therefore, two agents, one from offline training and one from online implementation, were compared and tested on one set of 10,000 initial states. Since the same set of initial states were used, the RL agent obtained after offline training was able to achieve a 7.91\% of failure rate which is the same as the value shown in Table~\ref{tbl:test_fail_rate}. The RL agent obtained after online training was able to reach a 0.02\% failure rate. By comparing these values, it shows that the proposed online implementation not only is able to ensures the stability, but also obtain a better RL agent. 
	
	{The closed-loop state and input trajectories of two episodes are plotted in Figures~\ref{fig:compare_episode_state_deter} and~\ref{fig:compare_episode_input_deter}, respectively. The markers on the figures are labeled with time instants, where $0$ represents the initial state ($x_0$), $200$ represents the final state ($x_f$), and $10-199$ are skipped for clarity. It is interesting to note that the RL agent is capable of reaching the same steady state from different initial states, even though the reward function is not explicitly designed for this purpose.
	\begin{figure}
		\centering
		\includegraphics[width=0.6\textwidth]{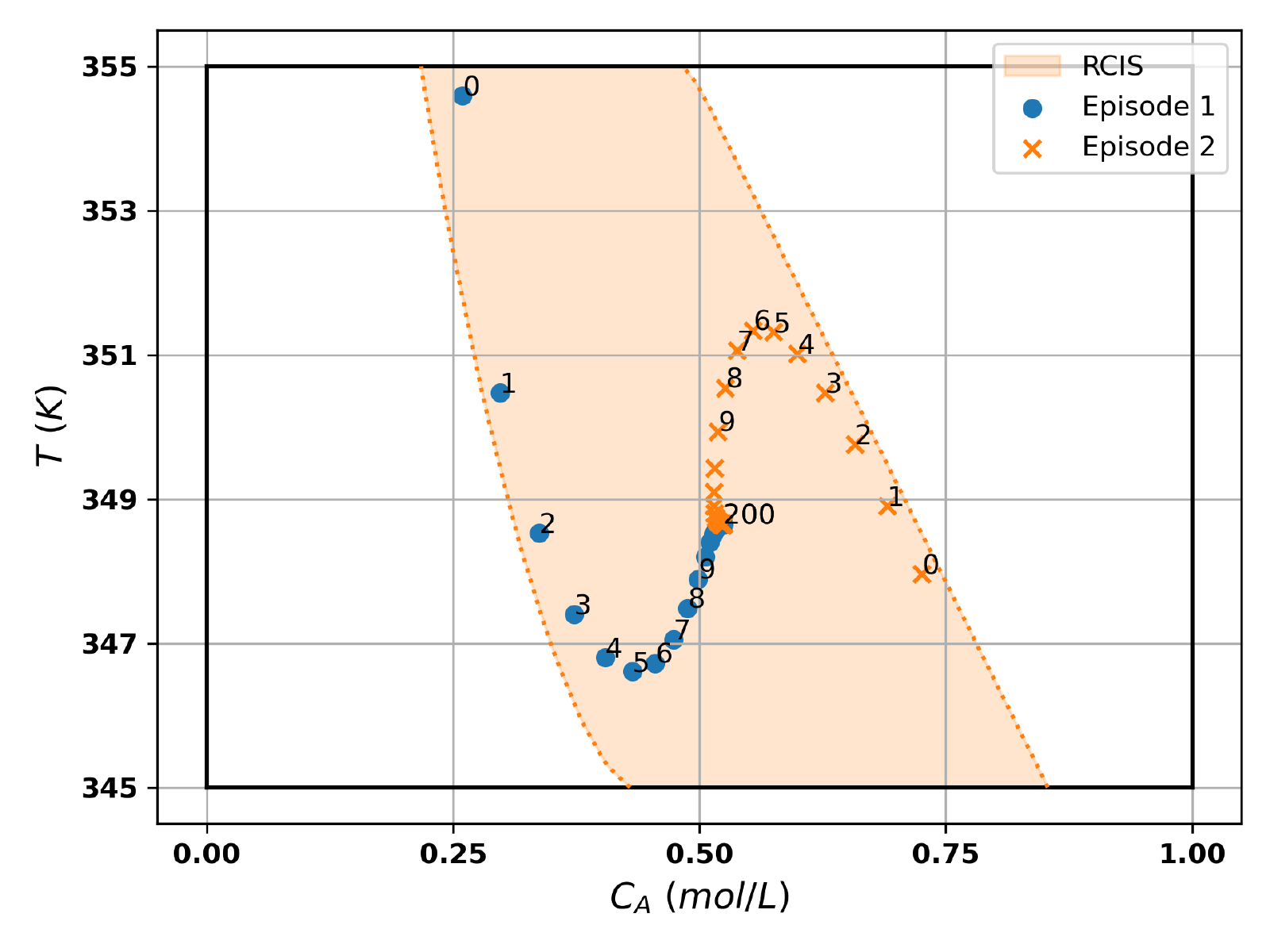}
		\caption{Two state trajectories of RL agents during online implementation.}
		\label{fig:compare_episode_state_deter}
	\end{figure}
	\begin{figure}
		\centering
		\includegraphics[width=0.6\textwidth]{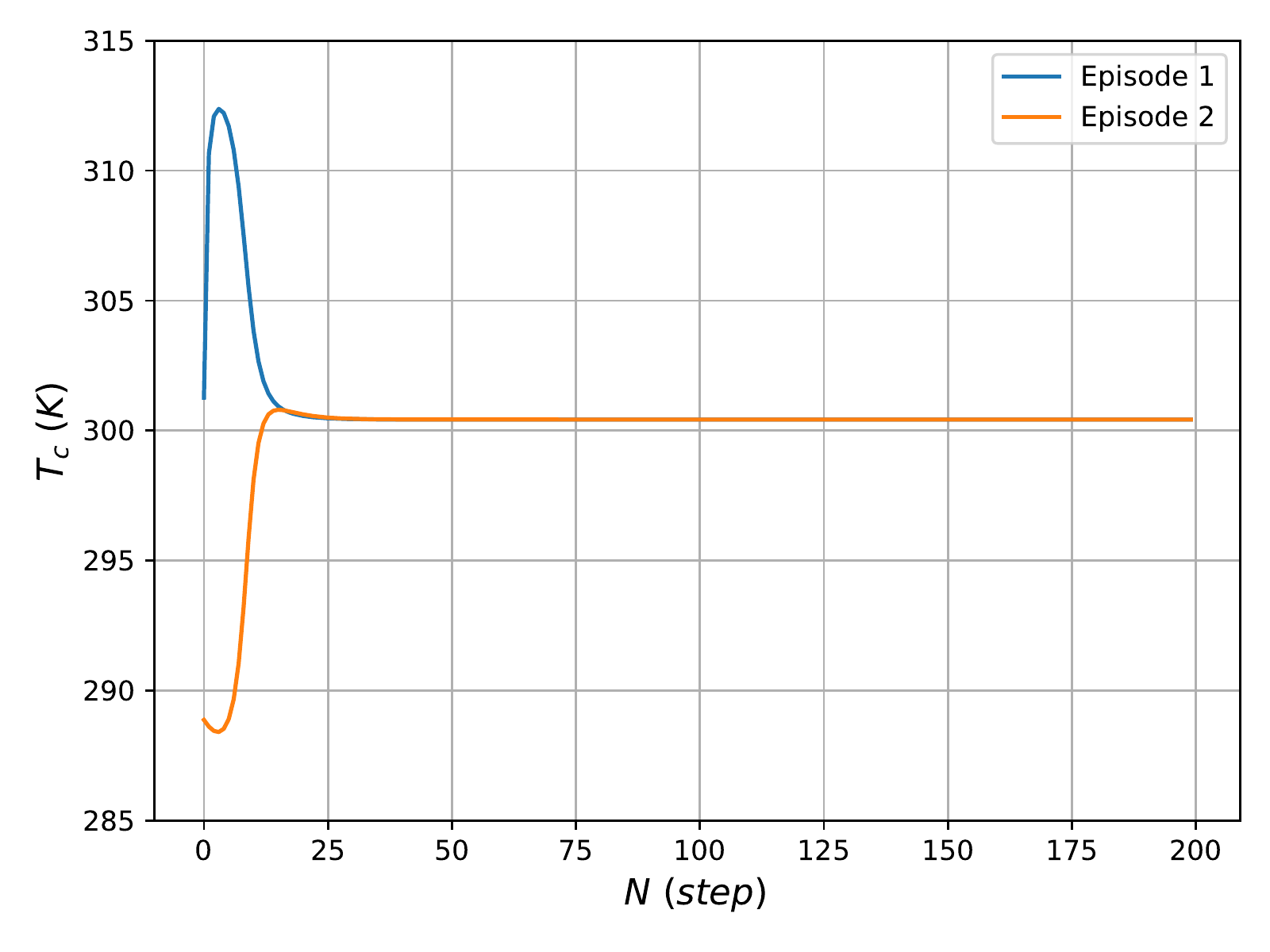}
		\caption{Two input trajectories of RL agents during online implementation.}
		\label{fig:compare_episode_input_deter} 
	\end{figure}
	
	{\section{Simulation results and discussion - with uncertainty}}
	In this section, the effectiveness of the proposed CIS-enhanced robust safe RL, as presented in Section~\ref{sec:proposed_stoc}, is evaluated in handling real-world scenarios with uncertainty. It is assessed from four perspectives: offline training with a focus on sampling efficiency, online stability guarantee, online time feasibility, and the incorporation of control objectives. Before delving into the experiment details and results, the RCIS was computed using the graph-based algorithm proposed in \cite{decardi2021computing}. The RCIS considered a bounded additive disturbance $|w|\leq [0.1,~2.0]^T$, which was also applied to the inlet concentration $c_{Af}$ and temperature $T_f$ of the CSTR. These variables, being generally influenced by upstream factors, can be measured but not manipulated. 
	Figure~\ref{fig:representation_of_rcis} illustrates the obtained RCIS, which is compared with the CIS in Figure~\ref{fig:representation_of_cis} and the physical constraints. The RCIS exhibits a similar shape to the CIS but with a smaller region due to the presence of uncertainty. Subsequently, the RCIS was utilized in both offline training and online implementation.
	
	\begin{figure}
		\centering
		\includegraphics[width=0.6\textwidth]{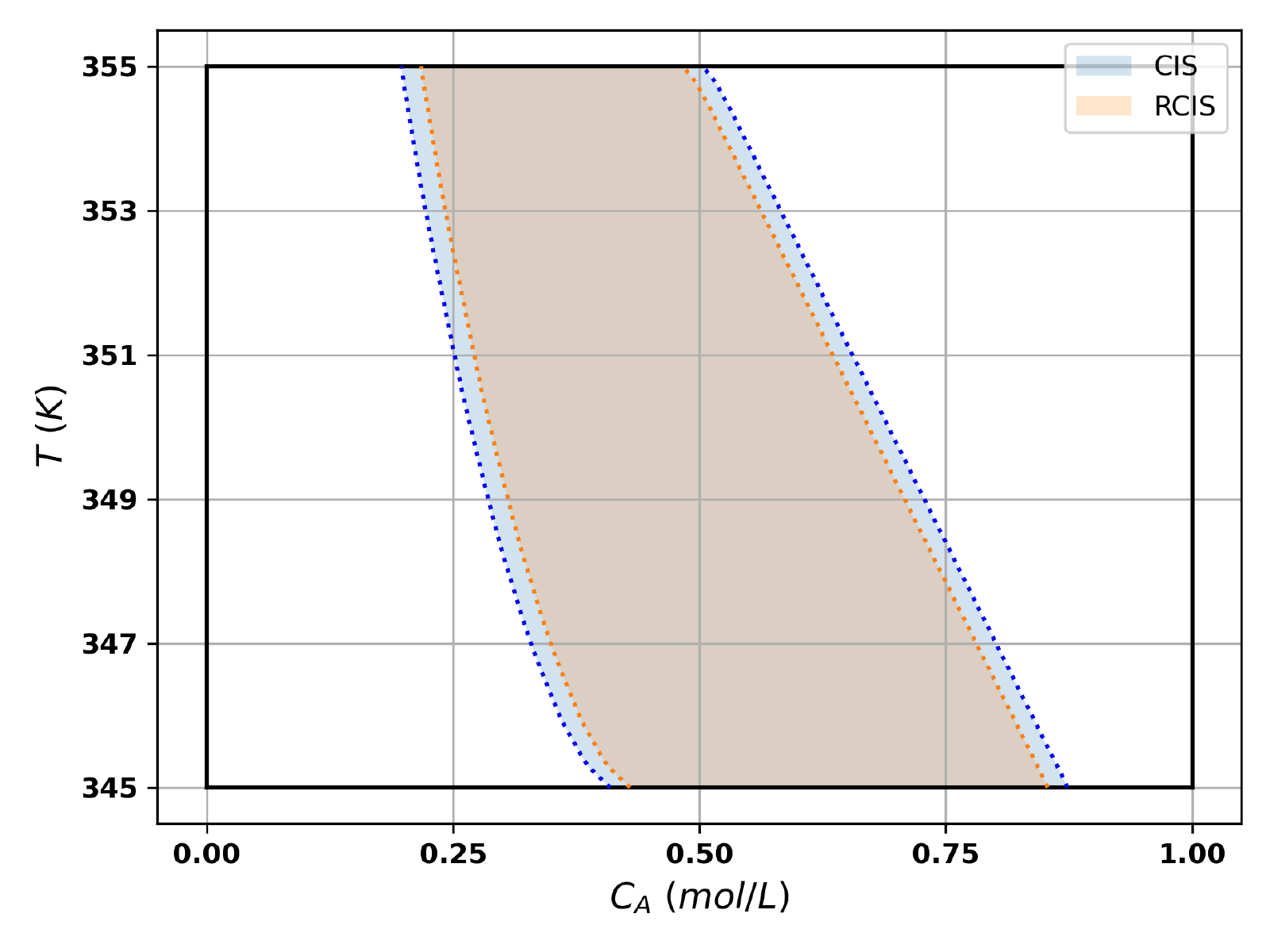}
		\caption{The physical constraints and the maximum CIS and RCIS of the CSTR.}
		\label{fig:representation_of_rcis} 
	\end{figure}

	\subsection{{Offline training and study of sampling efficiency}}\label{sec:rl_train_setup_stoc}
	Since the offline training algorithm remains the same for both the deterministic case in Section~\ref{sec:proposed_offline} and the one with uncertainty in Section~\ref{sec:proposed_stoc_offline}, the experimental setup follows the same procedure as described in Section~\ref{sec:rl_train_setup}. The only difference lies in the design of the reward function, which now depends on the RCIS. The new reward function was defined as follows:
	\begin{equation} \label{eqn:reward_stoc}
		r(x,u)= 
		\begin{cases}
			10,000& \text{if } \hat{x}_{k+1} \in R_r\\
			-1,000              & \text{otherwise}
		\end{cases}
	\end{equation}
	
	Similar to Figure~\ref{fig:train_reward}, the offline learning performance is assessed through the average training reward plot, as shown in Figure~\ref{fig:train_reward_stoc}.
	As the RL agent continues to interact with the environment over multiple episodes, we observe a clear improvement in its performance as indicated by the increasing episode scores. This signifies the agent's ability to effectively learn from the training process. In the initial phase, from the beginning up to around the {2,000th episode}, we observe a relatively higher learning rate accompanied by a larger variance among the 20 training runs. During this period, the agent is rapidly acquiring knowledge and adapting to the environment. However, as the training progresses, we notice a gradual slowdown in the learning rate, eventually reaching a plateau with a decreased variance. This indicates that the agent has reached a more stable state and has acquired a considerable level of proficiency in navigating the environment safely.
	
	\begin{figure}
		\centering
		\includegraphics[width=0.6\textwidth]{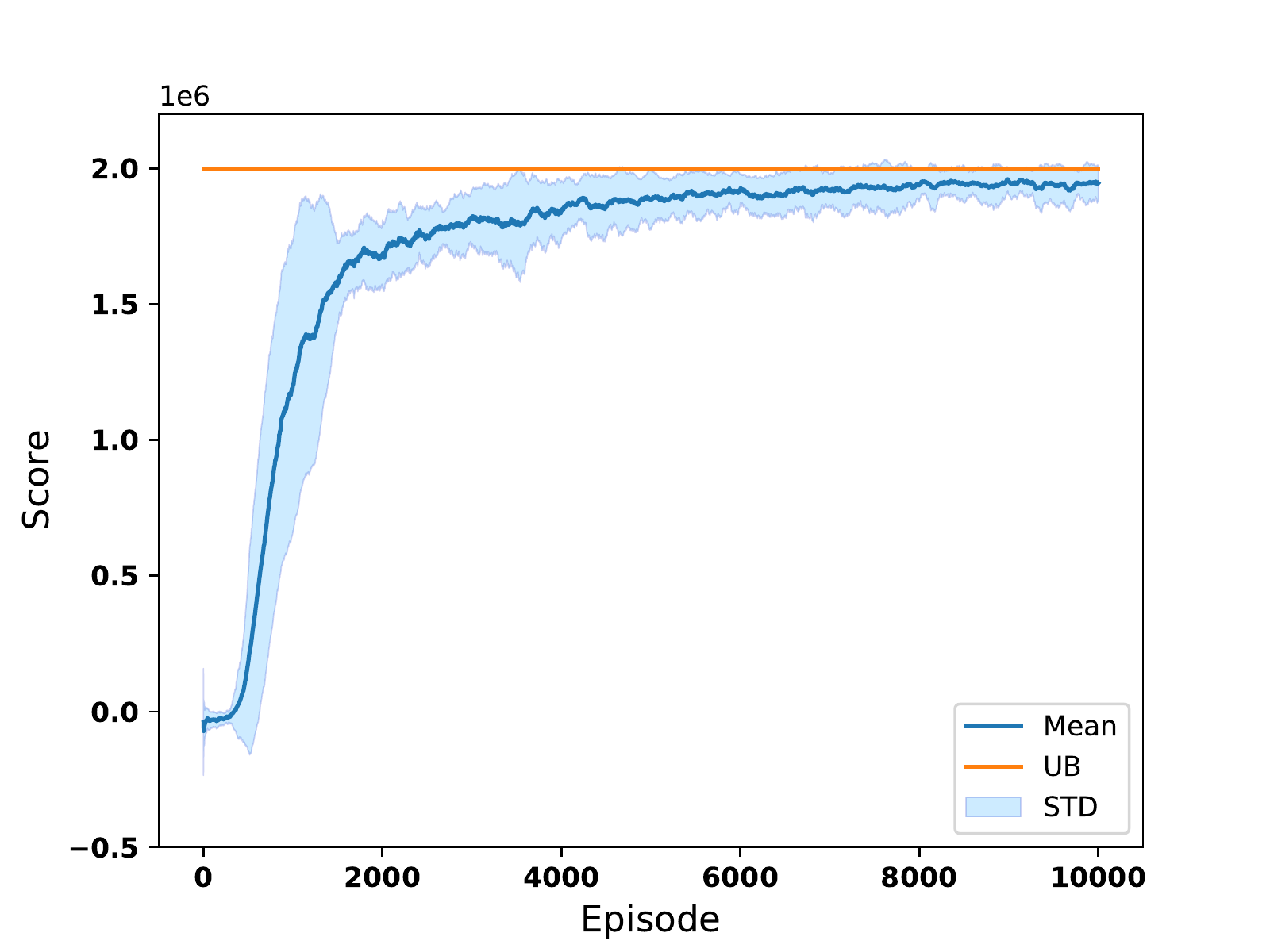}
		\caption{Average training score of proposed robust RL design}
		\label{fig:train_reward_stoc} 
	\end{figure}
	
	In terms of sampling efficiency, utilizing a training dataset of 10,000 samples, the RL agent with RCIS achieved a failure rate of 6.41\%, while the RL agent without RCIS had the same failure rate as the RL agent without CIS (34.30\%). Therefore, incorporating the RCIS improves the sampling efficiency, resulting in a lower failure rate during the training process.
	
	\subsection{{Study of online stability guarantee}}
	Upon completing the offline RL training using the RCIS, RL Run 1 was saved and utilized as a pre-calculated feedback controller for online implementation. The experimental setup remained the same as described in Section~\ref{sec:stability_guarantee}, with two notable differences. Firstly, uncertainty was introduced to the Environment block in Figure~\ref{fig:proposed_alg}. At each time step, when a safe action $u_k$ was applied to the Environment block, a sample of disturbance $w_k$ was uniformly drawn from the set $W$ and applied to the Environment block, resulting in a stochastic response from the process. Secondly, the MILP~\eqref{eqn:opt_milp}-\eqref{eqn:opt_milp_5} was solved in the Model block of the Safety Supervisor to verify the safety of the action $u_k$.
	
	The proposed online implementation strategy utilizing the MILP~\eqref{eqn:opt_milp}-\eqref{eqn:opt_milp_5} successfully kept the system within the RCIS for all 10,000 episodes. This outcome is expected, as the online implementation guarantees stability even in the presence of uncertainty. To illustrate the necessity and effectiveness of the MILP in the uncertain case, an additional experiment was conducted where RL Run 1 was implemented online without considering the worst-case uncertainty, and the Safety Supervisor solely relied on the constraint in~\eqref{eqn:polytope} to assess the safety of the action. The results showed a failure rate of 0.28\%, indicating that in 28 episodes, the predicted state $\hat{x}_{k+1}$ was within the RCIS, but the actual response $x_{k+1}$ was not.
	
	%\remove{Second, as the agent underwent retraining during the online implementation, it was anticipated that the retrained RL agent would exhibit improved stability performance. To evaluate this, we compared the performance of two agents: one obtained from the offline training phase and the other from the online implementation phase. Both agents were tested using the same set of 10,000 initial states. The RL agent obtained from the offline training phase achieved a failure rate of {\color{red}7.91\%}. In contrast, the RL agent obtained from the online implementation achieved an impressive failure rate of only {\color{red}0.02\%}. This comparison clearly demonstrates that the proposed online implementation not only guarantees stability but also results in an improved RL agent. {\color{red}How} can we guarantee that the agent after online training is always better than the agent after offline training?}
	
	\subsection{{Study of time usage}}
	To assess the computational feasibility of the proposed MILP optimization for ensuring the safety of the action $u_k$ during online implementation, its computational complexity was studied. Through simulations, it was determined that, on average, one MILP optimization required 0.20 seconds to solve. When the prescribed action $u_k$ was safe, no RL update was involved and only one MILP was solved. This computational time is acceptable within a sampling time of 6 seconds. In cases where the prescribed action was deemed unsafe, online retraining was conducted, and each update required 10 sets of data ($x_k$, $u_k$, ${r}_{k+1}$, $\hat{x}_{k+1}$). This resulted in the need to solve 10 MILPs per update, which took approximately 2 seconds. Considering a sampling time of 6 seconds, it is feasible to conduct 2 updates within this interval, ensuring the computational feasibility of online implementation.
	
	The above results were obtained using a computer with an Intel(R) Core(TM) i7-10700 CPU @ 2.90GHz processor and 16.0 GB RAM. The computational simulations were performed in Python 3.6, with the MILP modeled using Pyomo \cite{hart2011pyomo, bynum2021pyomo} and solved using Gurobi \cite{gurobi}.
	
	\subsection{{Study of control objectives}}
	In this section, we shift our focus from stability to addressing general control objectives, specifically economic control and economic zone control, as discussed in Section~\ref{sec:proposed_control_objectives}. Up until now, the RL agent was trained solely to keep the system within the invariant set, without considering other optimization goals. To incorporate these control objectives, we designed the following objective functions:
	
	\textbf{Economic maximization}
	\begin{equation} \label{eqn:reward_eco_stoc}
		r({x,u})= 
		\begin{cases}
			l_e(x)& \text{if } {\hat{x}_{k+1} \in R_r}\\
			-1,000              & \text{otherwise}
		\end{cases}
	\end{equation}
	\begin{equation}
		l_e(x) = 100(1-c_A)V
	\end{equation}
	where the economic performance is quantified by the mass of the reactant, which is the product of the concentration $c_A$ and the reactor volume $V$. The weight parameter of 100 allows for tuning the importance of economic performance. Additionally, the term $(1-c_A)$ is included to incentivize lower reactant mass when aiming for higher economic performance.
	
	\textbf{Economic zone tracking}
	\begin{equation} \label{eqn:reward_ecozone_stoc}
		r({x,u})= 
		\begin{cases}
			l_e(x) + l_z(x) & \text{if } {\hat{x}_{k+1} \in R_r}\\
			-1,000              & \text{otherwise}
		\end{cases}
	\end{equation}
	\begin{equation} \label{eqn:reward_zone_stoc}
		l_z(x)=\begin{cases}
			-300.0 \times (348.0-T)^2 &\textrm{for}~T<348.0\\
			0 &\textrm{for}~348.0\leq T\leq 352.0\\
			-300.0 \times (352.0-T)^2 &\textrm{for}~T>352.0\\
		\end{cases}
	\end{equation}
	where the reward function combines both economic optimization ($l_e(x)$) and zone tracking ($l_z(x)$). The zone tracking component is defined quadratically to penalize the distance between the state $T$ and the desired temperature zone, which ranges from 348.0 to 352.0. The weight parameter, with a magnitude of 300, emphasizes zone tracking over economic optimization when the state falls outside the target zone. The negative sign signifies that the agent incurs a penalty (subtraction from the reward) when it deviates from the target zone.
	
	The control objectives outlined above were incorporated into both the offline training and online implementation phases, with the aim of achieving economic control and economic zone control alongside stability maintenance. By following the experimental setup in Section~\ref{sec:rl_train_setup_stoc}, three RL agents were trained using different reward functions (equations~\eqref{eqn:reward_stoc},~\eqref{eqn:reward_eco_stoc}, and~\eqref{eqn:reward_ecozone_stoc}), which were subsequently implemented online using {500} initial states, each spanning 200 steps per episode. All three agents were able to maintain {500} episodes within the RCIS successfully. The economic performance for each online episode was evaluated using the following metric:
	\begin{equation}
		L_e = \sum_{k=1}^{200} 100(1-c_{A,k}V)
	\end{equation}
	By comparing the values of $L_e$ for the three RL agents across corresponding episodes that started from the same initial state, it was observed that the economic RL and economic zone tracking RL consistently outperformed the regular RL in terms of economic performance across all {500} episodes. This indicates that the inclusion of control objectives~\eqref{eqn:reward_eco_stoc} and~\eqref{eqn:reward_ecozone_stoc} yielded economic benefits. Table~\ref{tbl:eco_performance_3RLs} provides a summary of the average economic performance over {500} episodes for the three RL agents, further supporting the aforementioned observation.
	\begin{table}[t]
		\centering
		\renewcommand{\arraystretch}{1.5}
		\footnotesize
		\caption{Average economic performance for RLs trained with difference control objectives.}
		\begin{tabular}{ccc} 
			\hline
			Regular~\eqref{eqn:reward_stoc} & Economic~\eqref{eqn:reward_eco_stoc}  & Economic zone tracking~\eqref{eqn:reward_ecozone_stoc}\\
			\hline
			{964148} &      {985552} & {996982}\\  
			\hline
		\end{tabular}  \label{tbl:eco_performance_3RLs}
	\end{table}
	
	In order to evaluate the effectiveness of the zone tracking objective, we compared the closed-loop state trajectories of two RL agents: one trained solely with an economic objective and another trained with both an economic objective and a zone tracking objective. The trajectories are illustrated in Figure~\ref{fig:compare_eco_ecozone_state} and are generated from the same initial state. Both agents settles into their respective steady state neighborhoods. When comparing their economic performance without considering the zone violation, both agents achieved similar results, with the economic RL agent and economic zone tracking RL achieving 998,595 and 1,006,057, respectively, with a difference of 0.74\%. However, when we took into account the zone violation, we observed that the agent trained with zone tracking exhibited a more conservative behavior when deviating from the desired tracking zone, while the agent trained solely with the economic cost aimed to maximize economic performance.
	
	The final comment pertains to the optimality of the final steady state. An optimal solution to the steady-state optimization problem in equation~\eqref{eqn:ss_opt} yields the highest economic cost while satisfying the CIS constraint and corresponds to $x_s=[0.41,~354.98]^T$ and $u_s=298.68$.
	\begin{equation}\label{eqn:ss_opt}
		\begin{aligned}
			\max_{x_s, u_s} \quad & l_e(x_s)\\
			\textrm{s.t.} \quad & x_s - f(x_s, u_s, \textbf{0})=\textbf{0} \\
			\quad & Ax_s-b\leq \textbf{0}
		\end{aligned}
	\end{equation}
	However, the final steady state neighbors of two RL agents are located far from this optimal solution, suggesting that the RL agents prioritize safety over economic performance. Even though the optimal steady state, $x_s=[0.41,~354.98]^T$, is located close to the boundary, the RL agents may view it as a high risk of violating the bounds due to the balance between the control objective and safety in the reward function. Therefore, the RL agents' optimal solutions, which balance both economy and safety simultaneously, are located far away from the boundary.
	
	\begin{figure}
		\centering
		\includegraphics[width=0.6\textwidth]{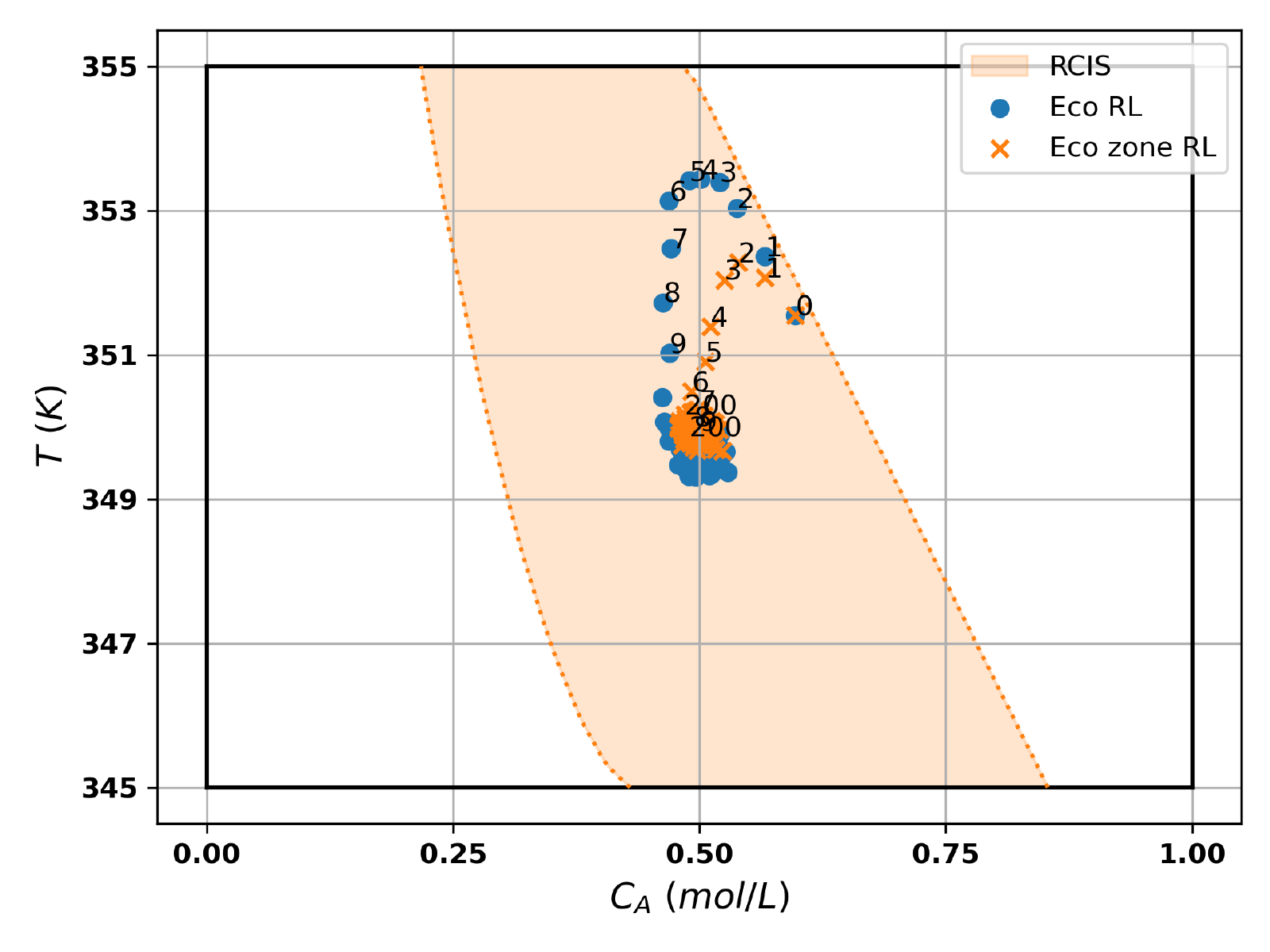}
		\caption{State trajectories of RL agents during online implementation. Blue dot: economic reward; orange cross: economic zone tracking}
		\label{fig:compare_eco_ecozone_state} 
	\end{figure}

\section{Conclusion}

A novel approach was proposed, combining control invariant set (CIS) with reinforcement learning (RL), to achieve stability-guaranteed RL implementation. The offline training stage integrated the CIS into reward function design, initial state sampling, and state reset technique. Additionally, an online implementation strategy was developed to ensure stability during RL execution, with the option for retraining when encountering new situations. Furthermore, the proposed algorithm was enhanced by incorporating Mixed Integer Linear Programming (MILP) to provide stability guarantees in the presence of uncertainty and stability proofs were provided for both deterministic and uncertain cases. The approach was applied to a two-dimensional nonlinear CSTR system, demonstrating improved performance compared to RL without CIS. The offline training stage reduced the failure rate and improved sampling efficiency. The online implementation stage ensured stability and resulted in a more effective RL agent in maintaining system stability within the CIS. Furthermore, economic and economic zone tracking control objectives were integrated into the algorithm, enabling optimization of both economy and safety. Time studies were conducted to assess the computational efficiency of solving MILP problems. Overall, the proposed approach offers a promising framework for achieving stability-guaranteed RL control while optimizing control objectives in complex systems.

\section{Acknowledgements}
Financial support from Natural Sciences and Engineering Research
Council of Canada and Alberta Innovates is gratefully acknowledged.

%\bibliographystyle{ieeetr}
%\bibliography{References}

\begin{thebibliography}{10}

\bibitem{mayne2000constrained}
D.~Q. Mayne, J.~B. Rawlings, C.~V. Rao, and P.~O. Scokaert, ``Constrained model
  predictive control: Stability and optimality,'' {\em Automatica}, vol.~36,
  no.~6, pp.~789--814, 2000.

\bibitem{sutton2018reinforcement}
R.~S. Sutton and A.~G. Barto, {\em Reinforcement learning: An introduction}.
\newblock MIT press, 2018.

\bibitem{mnih2015human}
V.~Mnih, K.~Kavukcuoglu, D.~Silver, A.~A. Rusu, J.~Veness, M.~G. Bellemare,
  A.~Graves, M.~Riedmiller, A.~K. Fidjeland, G.~Ostrovski, {\em et~al.},
  ``Human-level control through deep reinforcement learning,'' {\em Nature},
  vol.~518, no.~7540, pp.~529--533, 2015.

\bibitem{singh2022reinforcement}
V.~Singh, S.-S. Chen, M.~Singhania, B.~Nanavati, A.~Gupta, {\em et~al.}, ``How
  are reinforcement learning and deep learning algorithms used for big data
  based decision making in financial industries--a review and research
  agenda,'' {\em International Journal of Information Management Data
  Insights}, vol.~2, no.~2, p.~100094, 2022.

\bibitem{chen2022reinforcement}
X.~Chen, G.~Qu, Y.~Tang, S.~Low, and N.~Li, ``Reinforcement learning for
  selective key applications in power systems: Recent advances and future
  challenges,'' {\em IEEE Transactions on Smart Grid}, vol.~13, no.~4,
  pp.~2935--2958, 2022.

\bibitem{garcia2015comprehensive}
J.~Garc{\i}a and F.~Fern{\'a}ndez, ``A comprehensive survey on safe
  reinforcement learning,'' {\em Journal of Machine Learning Research},
  vol.~16, no.~1, pp.~1437--1480, 2015.

\bibitem{osinenko2022reinforcement}
P.~Osinenko, D.~Dobriborsci, and W.~Aumer, ``Reinforcement learning with
  guarantees: a review,'' {\em IFAC-PapersOnLine}, vol.~55, no.~15,
  pp.~123--128, 2022.

\bibitem{gu2022review}
S.~Gu, L.~Yang, Y.~Du, G.~Chen, F.~Walter, J.~Wang, Y.~Yang, and A.~Knoll, ``A
  review of safe reinforcement learning: Methods, theory and applications,''
  {\em arXiv preprint arXiv:2205.10330}, 2022.

\bibitem{kadota2006discounted}
Y.~Kadota, M.~Kurano, and M.~Yasuda, ``Discounted markov decision processes
  with utility constraints,'' {\em Computers \& Mathematics with Applications},
  vol.~51, no.~2, pp.~279--284, 2006.

\bibitem{chow2017risk}
Y.~Chow, M.~Ghavamzadeh, L.~Janson, and M.~Pavone, ``Risk-constrained
  reinforcement learning with percentile risk criteria,'' {\em The Journal of
  Machine Learning Research}, vol.~18, no.~1, pp.~6070--6120, 2017.

\bibitem{law2005risk}
E.~L. Law, ``Risk-directed exploration in reinforcement learning,'' 2005.

\bibitem{gehring2013smart}
C.~Gehring and D.~Precup, ``Smart exploration in reinforcement learning using
  absolute temporal difference errors,'' in {\em Proceedings of the 2013
  international conference on Autonomous agents and multi-agent systems},
  pp.~1037--1044, 2013.

\bibitem{zanon2019practical}
M.~Zanon, S.~Gros, and A.~Bemporad, ``Practical reinforcement learning of
  stabilizing economic mpc,'' in {\em 2019 18th European Control Conference
  (ECC)}, pp.~2258--2263, IEEE, 2019.

\bibitem{gros2021reinforcement}
S.~Gros and M.~Zanon, ``Reinforcement learning based on mpc and the stochastic
  policy gradient method,'' in {\em 2021 American Control Conference (ACC)},
  pp.~1947--1952, IEEE, 2021.

\bibitem{yu2018towards}
Y.~Yu, ``Towards sample efficient reinforcement learning.,'' in {\em IJCAI},
  pp.~5739--5743, 2018.

\bibitem{ma2021model}
H.~Ma, J.~Chen, S.~Eben, Z.~Lin, Y.~Guan, Y.~Ren, and S.~Zheng, ``Model-based
  constrained reinforcement learning using generalized control barrier
  function,'' in {\em 2021 IEEE/RSJ International Conference on Intelligent
  Robots and Systems (IROS)}, pp.~4552--4559, IEEE, 2021.

\bibitem{decardi2022robust}
B.~Decardi-Nelson and J.~Liu, ``Robust economic model predictive control with
  zone tracking,'' {\em Chemical Engineering Research and Design}, vol.~177,
  pp.~502--512, 2022.

\bibitem{blanchini1999set}
F.~Blanchini, ``Set invariance in control,'' {\em Automatica}, vol.~35, no.~11,
  pp.~1747--1767, 1999.

\bibitem{alshiekh2018safe}
M.~Alshiekh, R.~Bloem, R.~Ehlers, B.~K{\"o}nighofer, S.~Niekum, and U.~Topcu,
  ``Safe reinforcement learning via shielding,'' in {\em Proceedings of the
  AAAI Conference on Artificial Intelligence}, vol.~32, 2018.

\bibitem{gros2020safe}
S.~Gros, M.~Zanon, and A.~Bemporad, ``Safe reinforcement learning via
  projection on a safe set: How to achieve optimality?,'' {\em
  IFAC-PapersOnLine}, vol.~53, no.~2, pp.~8076--8081, 2020.

\bibitem{li2020robust}
S.~Li and O.~Bastani, ``Robust model predictive shielding for safe
  reinforcement learning with stochastic dynamics,'' in {\em 2020 IEEE
  International Conference on Robotics and Automation (ICRA)}, pp.~7166--7172,
  IEEE, 2020.

\bibitem{tabas2022computationally}
D.~Tabas and B.~Zhang, ``Computationally efficient safe reinforcement learning
  for power systems,'' in {\em 2022 American Control Conference (ACC)},
  pp.~3303--3310, IEEE, 2022.

\bibitem{brunke2022safe}
L.~Brunke, M.~Greeff, A.~W. Hall, Z.~Yuan, S.~Zhou, J.~Panerati, and A.~P.
  Schoellig, ``Safe learning in robotics: From learning-based control to safe
  reinforcement learning,'' {\em Annual Review of Control, Robotics, and
  Autonomous Systems}, vol.~5, pp.~411--444, 2022.

\bibitem{li2021comparison}
Z.~Li, ``Comparison between safety methods control barrier function vs.
  reachability analysis,'' {\em arXiv preprint arXiv:2106.13176}, 2021.

\bibitem{zhang2020}
Y.~Zhang, B.~Decardi-Nelson, J.~Liu, J.~Shen, and J.~Liu, ``Zone economic model
  predictive control of a coal-fired boiler-turbine generating system,'' {\em
  Chemical Engineering Research and Design}, vol.~153, pp.~246--256, 2020.

\bibitem{ellis2014}
M.~Ellis, H.~Durand, and P.~D. Christofides, ``A tutorial review of economic
  model predictive control methods,'' {\em Journal of Process Control},
  vol.~24, pp.~1156--1178, 2014.

\bibitem{rungger2017computing}
M.~Rungger and P.~Tabuada, ``Computing robust controlled invariant sets of
  linear systems,'' {\em IEEE Transactions on Automatic Control}, vol.~62,
  no.~7, pp.~3665--3670, 2017.

\bibitem{rakovic2005invariant}
S.~V. Rakovic, E.~C. Kerrigan, K.~I. Kouramas, and D.~Q. Mayne, ``Invariant
  approximations of the minimal robust positively invariant set,'' {\em IEEE
  Transactions on Automatic Control}, vol.~50, no.~3, pp.~406--410, 2005.

\bibitem{homer2017constrained}
T.~Homer and P.~Mhaskar, ``Constrained control lyapunov function-based control
  of nonlinear systems,'' {\em Systems \& Control Letters}, vol.~110,
  pp.~55--61, 2017.

\bibitem{fiacchini2010computation}
M.~Fiacchini, T.~Alamo, and E.~F. Camacho, ``On the computation of convex
  robust control invariant sets for nonlinear systems,'' {\em Automatica},
  vol.~46, no.~8, pp.~1334--1338, 2010.

\bibitem{decardi2021computing}
B.~Decardi-Nelson and J.~Liu, ``Computing robust control invariant sets of
  constrained nonlinear systems: A graph algorithm approach,'' {\em Computers
  \& Chemical Engineering}, vol.~145, p.~107177, 2021.

\bibitem{chen2021learning}
S.~Chen, M.~Fazlyab, M.~Morari, G.~J. Pappas, and V.~M. Preciado, ``Learning
  region of attraction for nonlinear systems,'' in {\em 2021 60th IEEE
  Conference on Decision and Control (CDC)}, pp.~6477--6484, IEEE, 2021.

\bibitem{bonzanini2022scalable}
A.~D. Bonzanini, J.~A. Paulson, G.~Makrygiorgos, and A.~Mesbah, ``Scalable
  estimation of invariant sets for mixed-integer nonlinear systems using active
  deep learning,'' in {\em 2022 IEEE 61st Conference on Decision and Control
  (CDC)}, pp.~3431--3437, IEEE, 2022.

\bibitem{bo2023control}
S.~Bo, X.~Yin, and J.~Liu, ``Control invariant set enhanced reinforcement
  learning for process control: improved sampling efficiency and guaranteed
  stability,'' in {\em IEEE Conference on Decision and Control}, submitted, 2023.

\bibitem{cannon2003nonlinear}
M.~Cannon, V.~Deshmukh, and B.~Kouvaritakis, ``Nonlinear model predictive
  control with polytopic invariant sets,'' {\em Automatica}, vol.~39, no.~8,
  pp.~1487--1494, 2003.

\bibitem{decardi2022graph}
B.~Decardi-Nelson, {\em Graph-based Computation of Control Invariant Sets:
  Algorithms, Analysis and Applications}.
\newblock PhD thesis, University of Alberta, 2022.

\bibitem{sioshansi2017optimization}
R.~Sioshansi, A.~J. Conejo, {\em et~al.}, ``Optimization in engineering,'' {\em
  Cham: Springer International Publishing}, vol.~120, 2017.

\bibitem{schulman2017proximal}
J.~Schulman, F.~Wolski, P.~Dhariwal, A.~Radford, and O.~Klimov, ``Proximal
  policy optimization algorithms,'' 2017.

\bibitem{hart2011pyomo}
W.~E. Hart, J.-P. Watson, and D.~L. Woodruff, ``Pyomo: modeling and solving
  mathematical programs in python,'' {\em Mathematical Programming
  Computation}, vol.~3, no.~3, pp.~219--260, 2011.

\bibitem{bynum2021pyomo}
M.~L. Bynum, G.~A. Hackebeil, W.~E. Hart, C.~D. Laird, B.~L. Nicholson, J.~D.
  Siirola, J.-P. Watson, and D.~L. Woodruff, {\em Pyomo--optimization modeling
  in python}, vol.~67.
\newblock Springer Science \& Business Media, third~ed., 2021.

\bibitem{gurobi}
{Gurobi Optimization, LLC}, ``{Gurobi Optimizer Reference Manual},'' 2023.

\end{thebibliography}
\end{document}